\RequirePackage[hyphens]{url}
\documentclass[a4paper,12pt]{article}

\usepackage{graphicx}
\usepackage{amssymb}
\usepackage{amsmath}
\usepackage{longtable}
\usepackage[dvipsnames]{xcolor}
\usepackage{float}
\usepackage{subfig}
\usepackage{caption}

\usepackage{cite}
\usepackage[margin=1.178in]{geometry}
\usepackage[affil-it]{authblk}

\usepackage{hyperref}
\hypersetup{
    colorlinks=true,
    urlcolor=blue
}
\usepackage{mathtools}

\usepackage[T1]{fontenc}  

\renewcommand{\it}{\normalfont \itshape }

\def\periodfl#1{\pflA#1.}
\def\pflA#1#2{\dot#1\ifx.#2\else\afterfi{\pflB#2}\fi}
\def\pflB#1#2{\ifx.#2\dot#1\else#1\afterfi{\pflB#2}\fi}

\begin{document}

\title{Distorted Sounds: Unlocking the Physics of Modern Music}

\author{Anna Mullin\thanks{Email: \texttt{anna.mullin@adelaide.edu.au}}\,\, 
    and Derek B. Leinweber\thanks{Email: \texttt{derek.leinweber@adelaide.edu.au}}} 
\affil{Special Research Centre for the Subatomic Structure of Matter, Department of Physics, The
  University of Adelaide, SA, 5005, Australia.} 
\maketitle
\begin{abstract}
In the production of modern music, the musical characteristics of the guitar or keyboard amplifier
play an integral role in the creative process. This article explores the physics of music with an
emphasis on the role of distortion in the amplification. In particular, we derive and illustrate how
a distorted amplifier creates new musical notes that are not played by the musician, greatly
simplifying the playing technique. 
In providing a comprehensive understanding, we commence with a discussion of the physics of music,
highlighting the harmonic series and its relation to pleasing harmonies. This is placed in the
context of the standard music notation of intervals and their relation to note frequency ratios.
We then discuss the problems of tuning an instrument and why the equal temperament of standard
guitar tuners is incompatible with good sounding music when amplifier distortion is involved.
Drawing on the basic trigonometric identities for angle sums and differences, we show how the
nonlinear amplification of a distorted amplifier, generates new notes not played by the musician.
Here the importance of setting your guitar tuner aside and using your ear to tune is emphasised.
We close with a discussion of how humans decipher musical notes and why some highly distorted guitar
chords give the impression of low notes that are not actually there.  
This article will be of assistance to students interested in the physics of music and lecturers
seeking fascinating and relevant applications of mathematical trigonometric relations and physics
to capture the attention of their students.
\end{abstract}

\newpage
\section{Introduction}

For the everyday music lover and the skilled musician alike, and for all who appreciate great
sounds – for the head-nodding, foot-tapping, back-of-the-pub ``I loved this song since forever!''
fanatics – music is transportive. 
As both physicists and musicians, we set out to understand the physics involved in producing modern
music, and we're surprised by the ways that frequencies are created and combined to produce musical
sounds. In some cases, the way in which those frequencies are produced is absolutely bizarre;
they're not played by the musician but are generated by the distorted guitar or keyboard amplifier
under use.
In understanding the physics of music, we find out why music appeals to so many of us. We learn how
frequencies create musical sounds, why some notes go together well and why it's almost impossible
to tune an instrument precisely.  We have also discovered how modern music benefits from surprising
effects in amplification distortion.

Let us first agree that we have no need to become music trivia champions or members of the symphony
orchestra to believe that great sounds are a pleasure. Similarly, we need no physics knowledge, and
yet so many physics phenomena are at the foundation of musical prowess; just as we need not
understand gravity to walk, musical physics often takes care of itself, both in the ways
frequencies interact when they are produced, and the ways that we process them in our technology
and in our brains. However, grasping better how those mechanisms benefit our music can allow us to
create and control richer sounds. Anyone frustrated with trying to tune a guitar will find great
peace of mind in understanding it's almost impossible.

We evaluate the likely mathematical reasons why music, though often obscure and confusing, sounds
good. For example, how do we enjoy both sweet harmonies by The Beatles and thundering metal by
Rammstein? Once we understand how to create a note from a fundamental frequency, we may then work
our way through layers of physical effects, which combine beautifully, from the instrument to the
chord and the tuning scale, and finally the surprising interaction with an amplifier. How much of
our response is subjective, how much is mathematical, and how much is both?

In the production of modern music, the musical characteristics of the guitar or keyboard amplifier
plays an integral role in the creative process. This article explores the physics of music with an
emphasis on the role of distortion in the amplification. In particular, we derive and illustrate how
a distorted amplifier creates new musical notes that are not played by the musician, greatly
simplifying the playing technique. 

In providing a comprehensive understanding, we commence in Sec.~\ref{sec:fundamentals} with a
discussion of the physics of music, highlighting the harmonic series and its relation to pleasing
harmonies. This is placed in the context of the standard music notation of intervals and their
relation to note frequency ratios.

In Sec.~\ref{sec:tuning} we discuss the problems of tuning an instrument and how this leads to a
modern Western scale with 12 notes.  We discover why the equal temperament of standard guitar tuners
is incompatible with good sounding music once amplifier distortion is involved.

In Sec.~\ref{sec:distortion}, we simulate how valve amplifiers distort the input signal and modify
the wave form. Two forms of nonlinear amplification are considered. Fourier transforms of the
amplified wave forms are calculated to show how the input frequencies are modified through the
process of distortion.  In particular, the creation of new tones is highlighted.

In Sec.~\ref{sec:trig}, we draw on the basic trigonometric identities for angle sums and
differences to explain how the nonlinear amplification of a distorted amplifier generates new
notes not played by the musician.  Here the importance of setting your guitar tuner aside and using
your ear to tune is emphasised.

Finally, in Sec.~\ref{sec:phantom} we close with a discussion of how humans decipher musical notes
and why some highly distorted guitar chords give the impression of low notes that are not actually
there.
This article will be of assistance to students interested in the physics of music and lecturers
seeking fascinating and relevant applications of mathematics and physics to capture the attention
of their audience.

\section{Fundamentals}
\label{sec:fundamentals}

\subsection{Complexity of a single note}

The first step towards explaining our responses to music comes from disentangling the complexity
hiding in a single note. The common note ``A3'', for example, is described as having a frequency of
220 Hz – a straightforward round number. However, this number is only one of many frequencies that
are involved in sounding out a real A3 note. When we play an A3 on an instrument, material inside
the instrument vibrates not just at 220 Hz, but also at integer multiples of that {\em fundamental
  frequency}, producing sound at 440 Hz, 660 Hz, 880 Hz, and so on. The combination acts as a
support network of harmonic tones, which aid the lowest tone by reinforcing it. The series of
higher harmonic vibrations round out and complete the sound, causing it to sound brighter. You can
experiment with using a sound visualiser to observe which frequencies are present in your own music
notes~\cite{inproceedings}.

These integer multiples of the fundamental, or {\em overtones} are amplified when they propagate
along a wind instrument's cavity or a piano's string because they are the only waves that can
resonate, having the correct length to fit the space. If they were to enter a game of Survivor
against all possible vibrations, they would champion almost instantly over the wrong-sized waves,
which are quickly exhausted by the counter-forces wearing them down in a space that does not
support their length. For a detailed account, check out historic books by American acoustician
Arthur H. Benade~\cite{BenadeInstruments}.  While the process of elimination may seem arbitrary, it
defines a profound measure of our appreciation of harmony: people enjoy listening to integer
multiples of one frequency, as our brains process all audible overtones at the same time. To
visualise the colossal amount of information hiding in simple sounds, try playing some audio
samples in an online spectrograph and watching the complexity unfold \cite{edwardball}. To see a
dramatic spectrogram revealing the frequencies involved with singing vibrato, take a look at this
video \cite{Michel_2017}.

\begin{figure}[tb]
\includegraphics[width=0.45\textwidth]{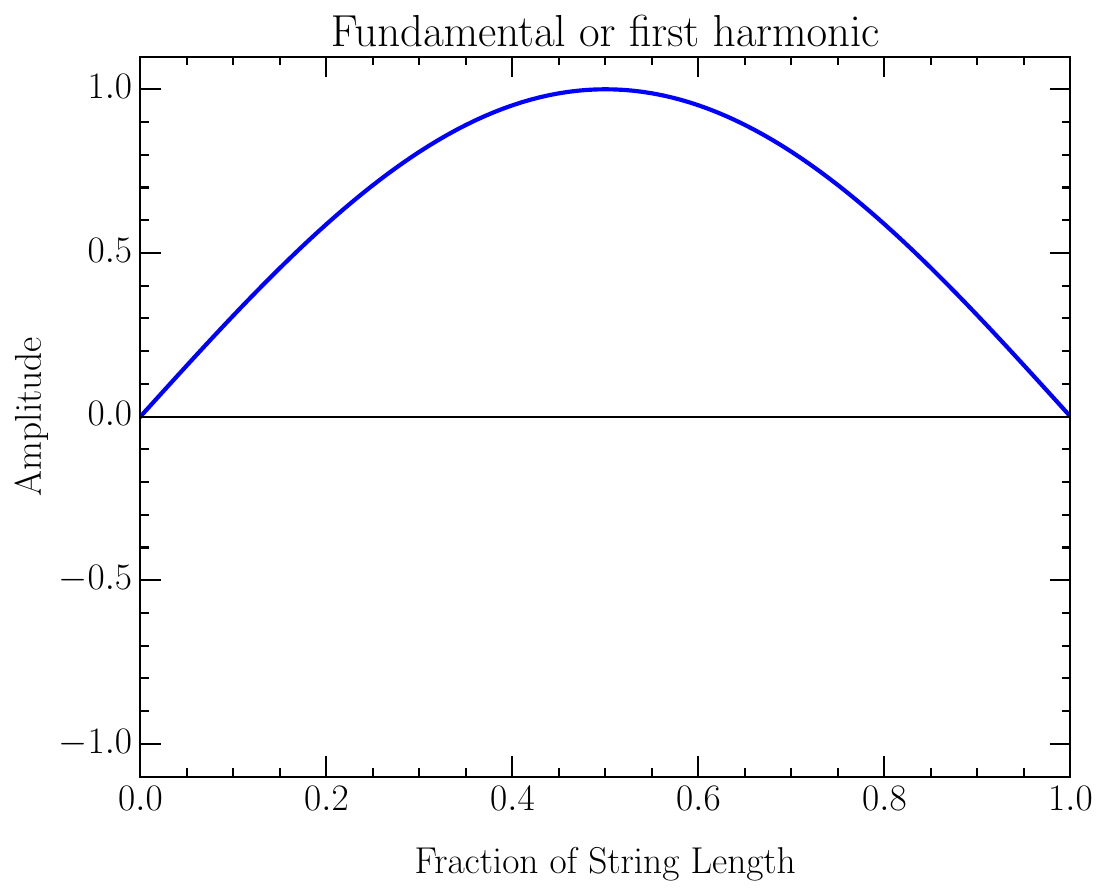}\qquad
\includegraphics[width=0.45\textwidth]{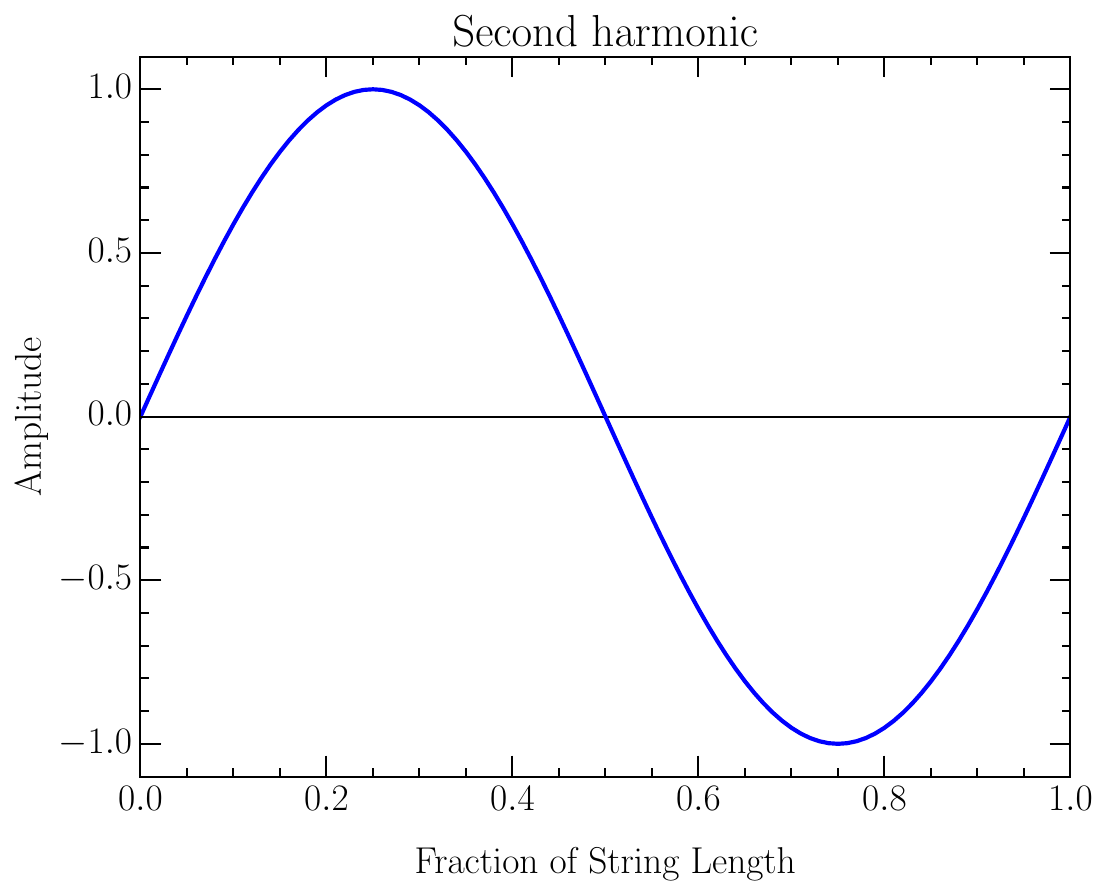}\\[6pt]
\includegraphics[width=0.45\textwidth]{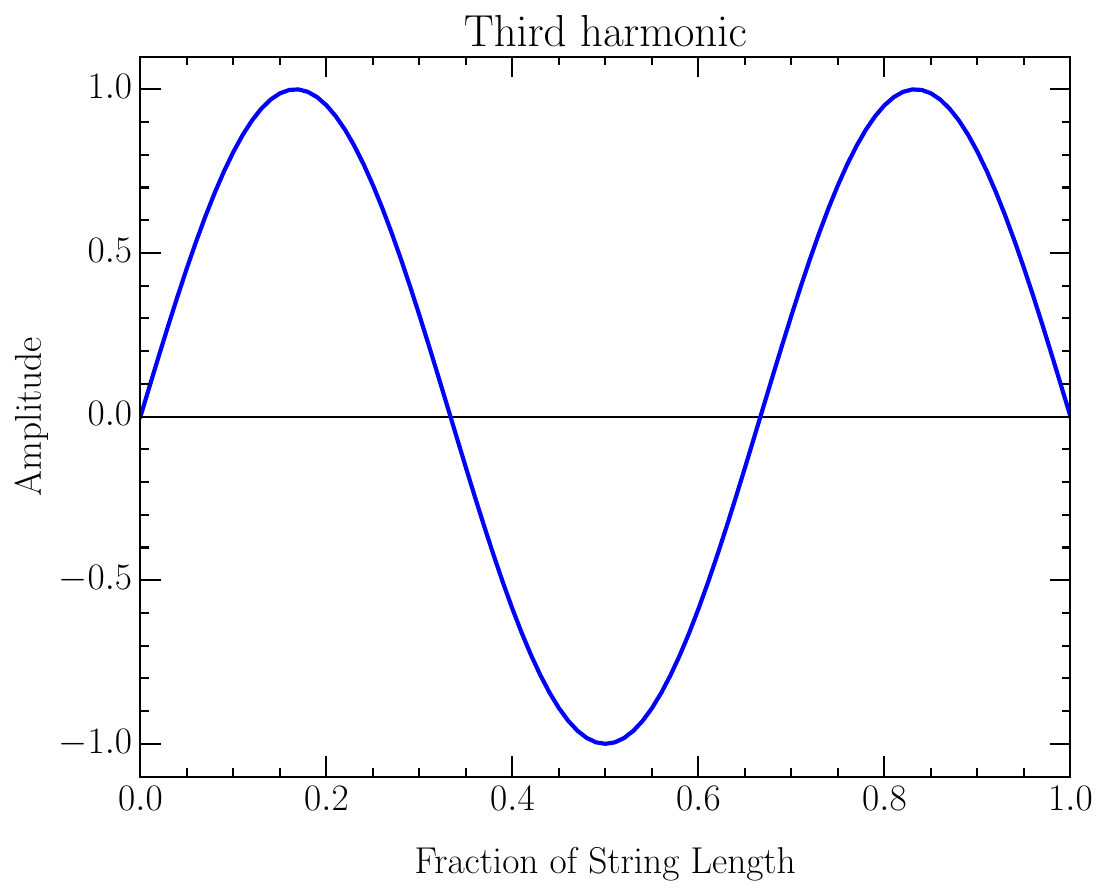}\qquad
\includegraphics[width=0.45\textwidth]{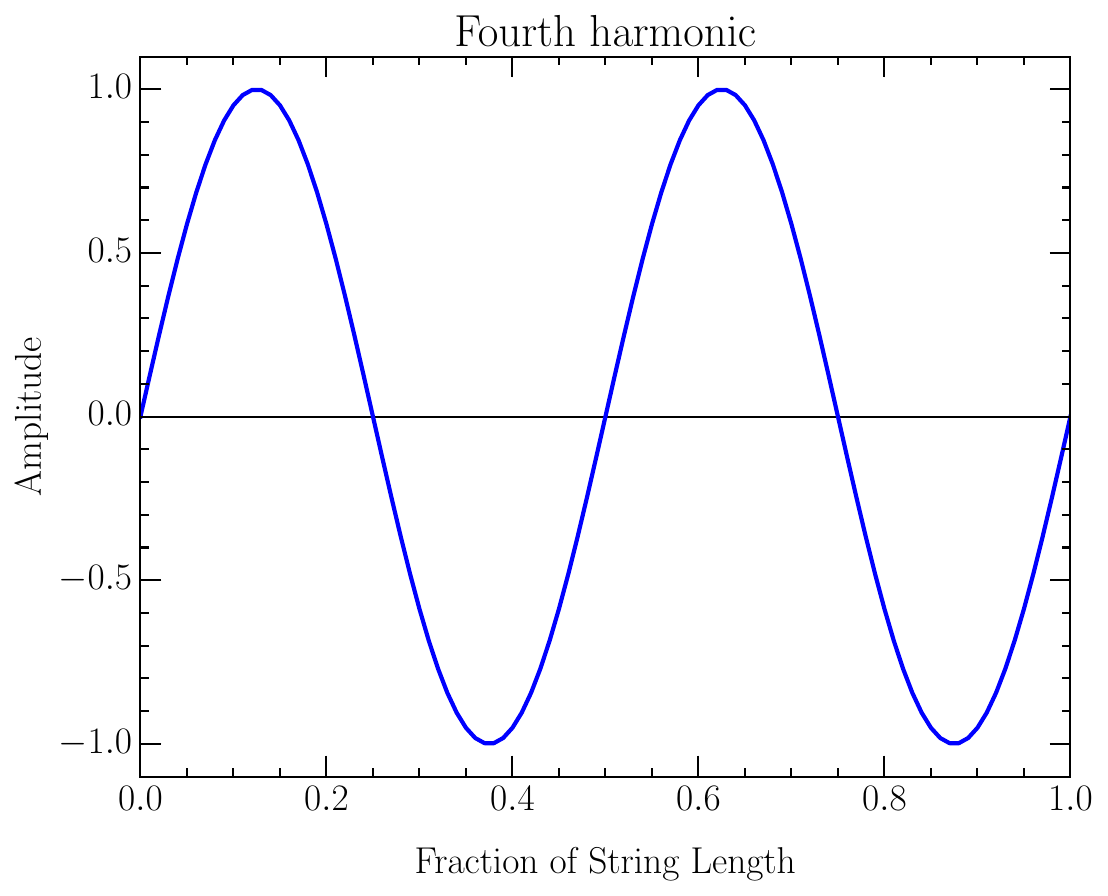}
\caption{The first four harmonic modes of vibration for an instrument string.}
\label{fig:wavemodes}
\end{figure}

The modes of vibration for a stringed instrument are illustrated in Fig.~\ref{fig:wavemodes}.
In this case, the waves must have zero amplitude at the two limits of the string.
The more waves that fit into the length, the higher the harmonic frequencies, and the brighter the
sound of the note.
If you pluck a guitar string or hammer a piano string, all these modes or harmonics are vibrating
at the same time.
And there are as many harmonics vibrating as there are numbers -- though only in theory. In
practice, harmonics above 10 are likely to be so faint that we cannot hear them.

The fundamental frequency and these overtones are referred to as the harmonic series of a single
note. The first harmonic is the fundamental frequency. We'll refer to each frequency in the series
as a tone.

\subsubsection{Mathematics in harmony}

Pythagoras -- yes, the one with the famous theorem -- noticed this phenomenon back in around 500
BCE when developing his theory that in music, order emerges from chaos in ways that we can predict
numerically. Today, we credit him as the originator of the belief that simple numerical proportions
bind musical parts together to form a perfect whole. The Pythagorean notion that we seek small
integer ratios is criticised for over-simplifying the musical experience, especially for neglecting
the perception, memory and recall of melody \cite{ParncuttRichard2018APTo}. Still, the Pythagorean
design of perfect harmony is essential to our understanding of pleasant sounds.

\subsubsection{Concept of timbre}

Although perfect harmonic integer multiples may still be important, they are not always our ideal
jam -- giving us further reason to question the Pythagoreans. Real instruments do not reproduce the
harmonic series exactly, because they contain tiny imperfections that change which vibrations
resonate best in the cavity or on the string. Leaving your guitar out in the summer sun might
encourage you to play it more, but the heat (and that thick layer of dust that has accumulated over
the past few months) are sure to change the sound, even after re-tuning. This happens because the
altered materials force the overtones to adopt slightly different frequencies and relative
amplitudes or loudness. Remarkably, we don’t need any equipment to discern the difference, because
our ears are highly efficient spectral analysers, recognising the precise sounds made by hundreds
of different frequencies at once.

Even before your guitar is warped in the hot sun, the set of overtones that it typically amplifies
are noticeably distinct from the overtones from an oboe, or a bassoon, or any other instrument, in
a phenomenon known as `timbre.’ \footnote{For a look at an often-cited cornerstone of
psychoacoustics and timbre, try the book by Helmholtz published as far back as 1885, in
Ref.~\cite{Helmholtz1885}. The book goes further to speculate about the role of our inner ears in
distinguishing timbres, which was groundbreaking at the time but in some cases corrected today.} It
is the overtones of the harmonic series that govern the timbre of a note played on an instrument.
While the pitch we perceive is that of the fundamental frequency, the character of the note is
governed by the loudness of each of the overtones and any slight deviations away from perfect
integer multiples of the fundamental frequency. An instrument's beautiful character comes from its
imperfections.

Real overtones are always a complex concoction of subtle amplitude and frequency imperfections
called inharmonicities. These combine in our ears and brains, and yet the timbre is still
recognisably distinct, although we often struggle to describe the differences without training. By
recognising each instrument’s trademark inharmonicity, we naturally have the power to identify the
material properties of different instruments from their sounds alone! If you wish to verify the
power of your auditory system, try using the sound visualiser again to check the difference between
overtones from the same note played on different instruments. You may even find out something new,
as we continue to make discoveries about timbre today. A stand-out (although slightly older)
example is the discovery that a banjo's characteristic bell-like twang comes from a unique
stretching of the plucked strings that is different from a guitar
\cite{politzer2014banjotimbrestringstretching}. The effect is easy to reproduce at home on free
sound editing software such as Audacity \cite{audacity_software}.

At the risk of offending Pythagoras, we might believe that listening to a range of imperfect
instruments produces a richer experience. Translating `rich’ into physics terms, we say that these
sounds contain `resonant frequencies close to integer multiples of the fundamental.’ Further
from the harmonic series, we perceive odder notes, like trying to identify an Australian accent in the
voice of a person who’s also spent years in Scotland, Spain and Uzbekistan – the complex
combination eventually becomes too confusing for us to identify a robust match. Overtones too far
from the harmonic series can sound confusing.
Indeed, a distinct timbre is not always created by inharmonicity, occurring also when the overtones
align with the original harmonic series but have different relative power.

\subsection{Harmony: Growing notes into chords}

Overtones that are close enough to the perfect harmonic series sound like a single note at the
fundamental frequency.  When the higher harmonics have larger amplitudes, the note sounds
brighter. But how does this concept of harmonics relate to the more standard use of the term
{\em harmony}? Harmony typically refers to consonant sounds between notes in a chord, where a chord
sandwiches together several series of frequencies at once, sometimes creating a delicious
concoction with pleasant ingredients (called consonance), and other times producing an
underwhelming cacophony of bizarre sounds (called dissonance). 

Usefully, we can use the integer multiples of the fundamental to determine which notes will combine
to produce the most delicious ``chord sandwich''. The recipe is this: consonant chords are composed
of notes with the most overlap between their overtone series \cite{SchoenbergA1983Toh}. Sound
tasty?

\begin{figure}[tb]
	\includegraphics[width=0.99\textwidth]{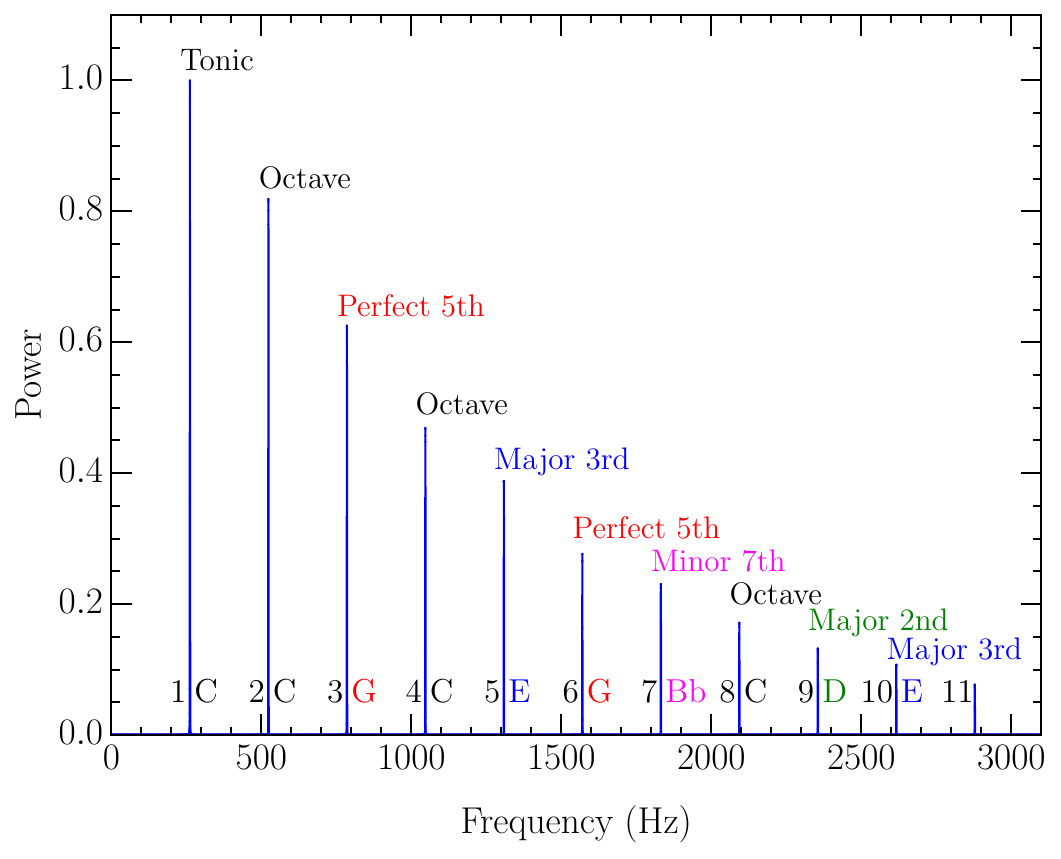}
	\caption{A power spectrum for a C4 note (an octave below middle C) with fundamental
          frequency 261.6 Hz played on a stringed instrument. Harmonics are labelled by harmonic
          numbers 1 through 11 and the note names for each tone in the series are indicated to the
          right of each peak. The first harmonic is the fundamental frequency of the note and the
          higher harmonics are integer multiples of the fundamental frequency.  The interval of
          each note from the fundamental is indicated above each peak. This nomenclature is
          reviewed in Sec.~\ref{sec:intervals}. }
	\label{fig:firstpowerspectrum}
\end{figure}

Using this idea as a guiding principle, we might decide to play a C note together with other notes
that have fundamental frequencies appearing in the C harmonic series, illustrated in
Fig.~\ref{fig:firstpowerspectrum}.\footnote{The frequency ratio of 11/8 relevant to the 11'th
harmonic is not a note in the 7-Limit Just tuning scale investigated in Sec.~\ref{sec:cantTune} and
therefore is not labelled by a note name.} Avoiding the harmonics that correspond with boring
repetitions of `C’ in different octaves, we can choose the third harmonic frequency, which happens
to be a G note, and the next novel frequency, an E note. Choosing any octave, if we play C, E and
G, then we find that surprisingly (or not!), this chord is beautifully consonant, a C-major
chord. Even so, a jarring B note might add a nice touch in a piece of music with a key of C major –
like a pinch of chilli to finish off the dish – even though it has fewer overlapping
harmonics. This is a C-major 7th chord, central to jazz music.

The choice of notes in a chord is the typical meaning of the word {\em harmony}, where aligning
overtones results in a consonant quality, and if misaligned they produce a cacophony of
dissonance. Although harsher, some dissonance is vital across genres, spanning classical and
modern, for fulfilling music.

The harmonic frequency series belonging to G and E notes above C are already contained in the C
series, so by playing a single C note, we effectively produce all three notes for the price of
one. To produce the harmonic C major triad more fully, we could enhance the E and G tone series
from amongst the existing frequencies in the C harmonic series, coaxing out this shy chord from
where it is hiding in plain sight. We just built a harmonic major triad by amplifying certain tones
that were already present in a single note! Let us break this down further.

\subsection{Musical Note Intervals}
\label{sec:intervals}

\begin{figure}[tb]
  \begin{center}
  \includegraphics[width=0.49\textwidth]{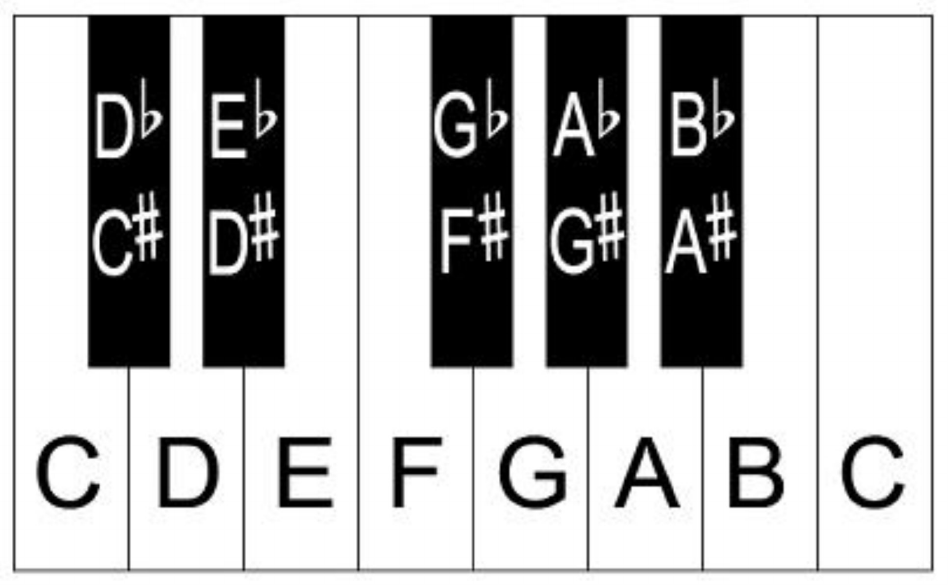}
  \hspace{1pt}
  \includegraphics[width=0.49\textwidth]{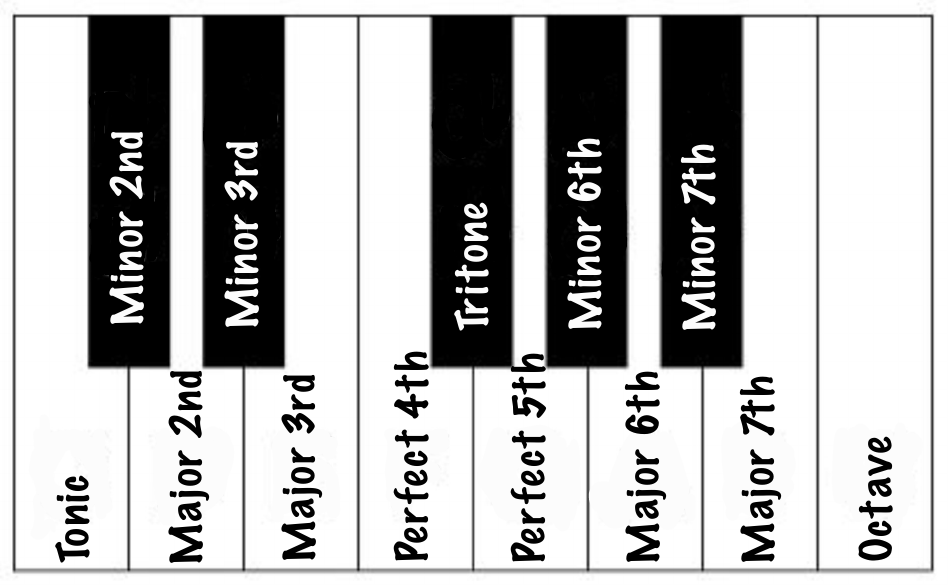}
  \end{center}	
  \caption{Piano keyboard illustrating (left) the names of the notes and (right) the names of the
    intervals for a scale in the key of C.  In the key of C, the positions of the black keys
    provide a pictorial guide to the names of the intervals for each note relative to the tonic.
    While the pictorial guide is specific to the key of C, the names of the intervals can be
    applied to any choice of note for the tonic.}
  \label{fig:pianoNotes}
\end{figure}

On a piano, the notes of major and minor scales in 12-tone music are laid out in patterns such as
the one in Figure \ref{fig:pianoNotes} for the key of C. Although there are 12 notes in this scale,
only 8 of them are attributed to the major scale, which drives the nomenclature. In the key of C,
all the notes of the C major scale are on the white keys of the piano. Counting the first note –
the tonic – as 1, we count 2 through 8 for the intervals belonging to the major scale, although the
8th is not a unique note to the tonic.

Thus, an interval is a difference in pitch between two sounds. It is best expressed as a ratio of
the fundamental frequencies of the two notes comprising the interval. As humans, we enjoy the
harmony of frequencies related by ratios of small integers.  In Western music, the first note of a
scale is called the tonic, and the frequency ratio:
\begin{itemize}
\addtolength{\itemsep}{-6pt}
\item 2/1 is an octave above the tonic,
\item 3/2 is the perfect fifth interval,
\item 4/3 is the perfect fourth interval,
\item 5/3 is frequency ratio of a major third interval.
\end{itemize}

The most harmonic interval, aside from the octave, is the third tone of the harmonic series of a
note, because this is the second-lowest and most dominant overtone. For a C note, the third
harmonic has the frequency of a G note, which is the perfect fifth interval as illustrated in
Fig.~\ref{fig:firstpowerspectrum}.
Next to the octave, the perfect fifth is associated with the ratio of the next smallest integers,
3/2. To see what these intervals look like as interfering waves, check out the YouTube video in
Ref.~\cite{Neron_2014}.

It is easy to confuse terminology, where the {\em third harmonic} and the {\em perfect fifth}
interval are closely related. Playing the two notes C and G together, we obtain the interval called
the perfect fifth. This is because the two notes have fundamental frequencies at the first note of
the C scale, called the tonic, and the fifth note of the C major scale.
The strong overlap between their harmonic series produces a powerful, consonant sound. 

The next-most-harmonic interval involves the fifth harmonic of the C harmonic series, an E note as
illustrated in Fig.~\ref{fig:firstpowerspectrum}. This is the fourth lowest overtone of the C
harmonic series. We are counting only the smallest integers which are unique, rather than repeats
of the tonic at higher octaves, and therefore discount the 2nd and 4th harmonics. The E note is the
second new note of the harmonic series, the {\em major third} of the C-major scale, and also
creates nice consonant overlap in harmonic series when played with the C note.

For a neat way to visualise the harmonic relationships in music, try the Tonnetz
\cite{tonnetzviz}. The method has the appearance of a lattice of tones, connected to each other by
the third or fifth interval. You can try charting chord progressions and watch how they are related
as they spin across the lattice in fascinating patterns.

\section{Tuning an instrument}
\label{sec:tuning}

Music can seem full of choices made long ago by history, and when we decide to pick up an
instrument today, these old choices leave us with an exciting legacy to unpack. Maybe they are just
arbitrary choices, or maybe they are mathematically motivated. For example, the beauty of the
perfect fifth is unquestionably a mathematical one, appearing as the lowest non-octave integer
multiple in a note's harmonic spectrum. We might also question why there are twelve notes in the
scales of Western music -- could this be an arbitrary choice? Actually, the idea was based on the
power of the perfect fifth, and is again explained by mathematics.

\subsection{Measuring the accuracy of tuning}

\subsubsection{Equal Temperament Scale}

We've seen how the most consonant perfect tuning involves ratios of frequencies that correspond to
ratios of small integers. The origin of these ratios lies in the harmonic series of a vibrating
string. Recall, the most consonant interval, a perfect fifth, corresponds to a ratio of the
fifth note of the major scale to the tonic note of the major scale, with a frequency ratio equal to 3/2.

However we find a problem to tarnish this perfection: we cannot tune more than one scale perfectly at a time.
Imagine retuning a piano just because the singer decides that their voice sounds better when the song is transposed to another scale. By the time you wait around for the tuning to be finished, everyone has gone home. 

To enable the consideration of any scale, compromises need to be made.  One needs the frequency ratio
between each pair of adjacent notes to be constant.
The process called {\em tempering the scale} is to make small adjustments, and thus compromises, to
each of the scale's notes. 

Equal temperament has equal multipliers between the frequencies of adjacent notes. It's usually
defined with reference to the note A4 having the exact fundamental frequency of 440 Hz with all
other note frequencies derived relative to A4.

Suppose our octave is divided into $N$ notes.  Then the equal-tempered interval between each of
these notes is a number which when raised to the $N$'th power (i.e. multiplied by itself $N$ times)
equals 2 for the octave.  Mathematically, this is the $N$'th root of 2.  Our standard chromatic
scale has $N = 12$.  The 12'th root of 2 is $2^{1/12} = 1.059463094\ldots$.  This is an irrational
number and thus differs from the rational frequency intervals of harmonic tuning.  In other words,
\begin{center}
	\noindent
	\framebox[0.95\textwidth]{
		\vspace*{3mm}
		\begin{minipage}{0.9\textwidth}
			\vspace*{3pt}
			\centerline{\normalfont \bfseries Equal temperament is always out of tune.}
			\vspace*{1pt}
		\end{minipage}
		\vspace*{3mm}}
\end{center}
\vspace{6pt}
This can cause big problems in modern music.
Hope you didn't pay too much for that guitar tuner! Trust us -- it's not what you want.

Fortunately, our $N = 12$ equal tempered scale closely matches the important harmonic tuning
intervals of the 4th, and 5th; they are off by a mere 2 one-hundredth's of the interval between
notes on a piano.  This is good news, as these intervals carry the team in harmonic value, so they
need to be as close as possible to perfect.
This is yet another reason why Western music has settled on 12 notes in the scale, adopted
in the early 1900's.
For a number that seems quite arbitrary at first glance, 12 is a powerful
number of notes to add to your scale.
With equal temperament, any note can be declared the tonic of the scale and the tuning of the scale
is equally acceptable.
So, even though equal temperament is out of tune, its attempt at harmonicity is worthy of
celebrating. After all, perhaps we shouldn't expect a system so useful to be perfect.

This deviation of 2 one-hundredth's of the interval between notes on a piano is referred to as a
tuning error of 2 {\em cents}.  We define the concept of cents mathematically in the next section.

\subsubsection{Measuring tuning deviations in cents}

The cent is a unit used in tuning to measure fine differences between the sound of an
instrument and the intended scale \cite{AngusJamie2009A3Cb}.
The cent is clearly named after the number 100, and we place one hundred cents between every note
of the equal-tempered scale, so a note 50 cents away from one of the equal tempered notes is the
most out of tune you can get.

With 12 notes in the Western scale, the frequency ratio between the $n$'th note in the equal
tempered scale and the tonic looks like this:
\[
\frac{f_n}{f_1} = 2^{(n-1)/12} = \left ( 2^{1/12} \right )^{(n-1)} 
                               = \left ( \sqrt[12]{2} \right )^{(n-1)} \, ,
\]
where the thirteenth note of the scale (the octave) has $n=13$, and 
\[
\frac{f_{13}}{f_1} = 2^{(13-1)/12} = 2^{12/12} = 2 \, ,
\]
a doubling of the frequency.  The concept of the cent is introduced by considering the frequency ratio
of the $n$'th note, $f_n$, relative to the tonic, $f_1$, as
\[
\frac{f_n}{f_1} = 2^{(n-1)/12} \equiv (n-1) \times 100\ \mbox{cents} = c\ \mbox{cents.}
\]
The first note, the tonic, is at $0$ cents, the octave is at $1200$ cents, and each note in between
is a multiple of one hundred cents.

For an arbitrary frequency ratio, $f/f_1$, one can solve for $n-1$ and then multiply it by 100 to
get the cents value. This time, $n$ will not necessarily be an integer.
\[
\frac{f}{f_1} = 2^{(n-1)/12} \, , \to \left ( \frac{f}{f_1} \right )^{12} = 2^{n-1}\, ,
\]
and taking the $\log$ of both sides
\[
12\, \log \frac{f}{f_1} = (n-1)\, \log 2\, \to n-1 = \frac{12}{\log 2}\, \log \frac{f}{f_1} \, .
\] 
recalling, the number of cents, $c$, is $(n-1) \times 100$
\[
c = \frac{1200}{\log 2}\, \log \frac{f}{f_1} \,.
\]
As expected, a doubling of the tonic frequency sets $f/f_1 = 2$ and $c = 1200$ for the octave.

A quirk of our auditory system is that between our ears and brains, we process frequencies as
ratios. We listen mostly to the relative difference in pitch, and those of us without perfect pitch
are challenged to identify absolute frequencies. The beauty of the measure of cents comes from how
well-suited this metric is to our biology and brains, since it produces an equal multiplicative
factor between the two frequencies of each consecutive pair of notes. We are holding this simple
factor $\sqrt[12]{2}$ constant between every consecutive pair, so we hear those notes as equally
spaced. Our concern for ratios is reflected on the logarithmic scale, where the notes become
equally spaced, the way we hear them. Thus, the measure of cents really is intuitive! We will later
use this concept to define the harmonic quality of a musical scale.

\subsection{12 notes}

Have you ever wondered why there are 12 notes in the musical scale of Western music?  It is not an
artistic choice, but a mathematical one that we trace back to Pythagoras in sixth century BCE. We
return to the Pythagorean ratios of small integers; the idea for 12 notes was to create a scale
based solely on the most harmonious ratio of frequencies equal to 3/2. In modern times, this ratio
of frequencies corresponds to the perfect 5th interval between notes, argued to be the most
consonant or pleasant to the ear.  It is the first overtone of the harmonic series that is not an
octave of the fundamental.  It is also one of the easiest intervals to tune by ear.

The scale is derived by choosing a frequency for the first note of the scale called the tonic. All
other notes are derived by either moving up by a fifth by multiplying the frequency by 3/2, or down
by multiplying the frequency by 2/3. As the new notes are generated, they can be brought
back to the octave of the tonic by dividing or multiplying by 2 as necessary.  This process of
``Pythagorean tuning" was used by musicians up to the beginning of the 16th century.  In modern
language this method of tuning is referred to as three-limit just (or harmonic) tuning.

After 6 steps up and 6 steps down, the two notes produced are almost the same note. They differ by
less than a quarter of a step between the keys of a piano.  After discarding one of the two notes,
one has a 12-tone scale, similar to today's equal temperament.  And for harmonies of the perfect
5th, the scale is wonderfully smooth and consonant.

To understand the special quality of the 12 tone scale, we can consider other numbers of tones in a
Pythagorean scale based on the most consonant interval of the fifth. The idea is to examine how far
the $N+1$'th note of an $N$-note scale is away from a perfect octave.

We can calculate how well a scale cycles to converge to its tonic by following some simple steps.  
\begin{enumerate}
\item Begin with the tonic frequency ratio, tonic:tonic, of 1.
\item Increase the current ratio by a perfect 5th by multiplying it by 3/2.
\item If this frequency ratio exceeds 4/3, then divide it by 2 to bring the interval back to the
  octave surrounding the tonic.\footnote{Note the limit of 4/3 is the boundary where dividing the
  frequency ratio by 2 leaves the ratio the same distance from unison.  To see this mathematically,
  denote the frequency ratio by $r_f$ and solve $r_f -1 = 1-r_f/2$ for $r_f$. Here the value of 1
  denotes the unison frequency ratio we aim to be close to.}

\item Now compare this frequency ratio with the unison ratio of 1 to see if the $N+1$'th note in
  the $N$ note scale is close to an octave. 

\item Next, consider adding another note to the scale by looping back to step 2 above.
\end{enumerate}

Table \ref{tab:numNotes} provides a summary of this algorithm by writing entries where the $N+1$'th
note in the scale is close to the tonic.  The table includes cases where the note produced in the
algorithm is within a tone of the standard 12-note equal-tempered musical scale.

\begin{table}[tb]
\caption{Results from the consideration of other numbers of notes in a musical scale up to 50
  notes.  When the number of notes in the Pythagorean scale produces an octave close to the tonic
  frequency multiplied by a power of 2, an entry is written to the table.  The differences from
  unison are quoted in cents, where 100 cents describes the distance between adjacent notes in the
  equal temperament scale of a piano.}
\label{tab:numNotes}
\centering
\begin{tabular}{ccccc}
\hline
\noalign{\smallskip}
Number of       &Approximate Unison   &Difference    &Within      \\
Notes in Scale  &Frequency Ratio      &from Unison   &Half a Tone \\
\noalign{\smallskip}
\hline
\hline
\noalign{\smallskip}
   5   &0.94922    & -90.2               \\
  12   &1.01364    &  23.5   &\checkmark \\
  17   &0.96217    & -66.8               \\
  24   &1.02747    &  46.9   &\checkmark \\
  29   &0.97530    & -43.3   &\checkmark \\
  36   &1.04149    &  70.4               \\
  41   &0.98860    & -19.8   &\checkmark \\
  48   &1.05570    &  93.8               \\
\noalign{\smallskip}
\hline
\end{tabular}
\end{table}

One can see that 12 notes is a very special case, where an octave returns to unison within 24
cents. There are 100 cents between adjacent notes, so just 24 cents is remarkably good. The only
case that comes closer within the first 50 notes is choosing 41 notes instead of 12, but that
leaves us with an overwhelming number of notes.

Although the 12-note choice is historically driven, in equal temperament tuning, any number of
notes is possible. Other numbers of notes have gained prominence, including 31, which lends
excellent approximations to the harmonic 3rd and harmonic 7th of the major scale, far surpassing
our best efforts in 12-tone equal temperament where the 7th is more than 30\% off. The 3rd features
prominently in modern music and the 7th is responsible for that unique barbershop sound.  This is
the tricky balancing act between creating a scale that's easy to use and one that sounds good.

\subsection{Harmonic quality of a musical scale} 

Now that we are convinced that we would like twelve notes in our musical scale, we are still left
to decide which frequencies to use, especially for modern music with lots of distortion. Ideal
ratios are clearly important, so let us start there; ratios of small rational integers, just like
the ones we found in the harmonic series, should help us to create our perfect harmonic scale.

Provided it's not too hard to re-tune an instrument ({\it e.g.} a guitar), this is a fail-safe way
to construct a scale. Most of the chords we choose to play will naturally sound pleasant. Plus,
with the competition in the annual Christmas choir concert getting tougher every year, we cannot
afford to settle for any less than ideal harmonic ratios. Such a scale allows us to sing a rousing
rendition of perfect intervals with no need for accompaniment -- and to impress even the most
discerning grandmothers in the audience.

Especially impressive is that power chord note that generates the perfect fifth interval. Starting
with the third harmonic frequency and dividing by two, we return to the first octave, giving us the
celebrated frequency for the perfect 5th of our scale, at $3/2$ times the frequency of the tonic.

Similarly, we can move down by multiplying the tonic by $2/3$ and then doubling the frequency to
arrive back in the octave. This is the perfect 4th of our scale, with a frequency of $4/3$ times
the tonic frequency. According to our brains, this ratio is small enough to sound nice and
harmonic, so it should make the cut too.

However, there is a problem with this approach. The intervals between these notes are not
uniform. We will need to tune a new scale for every different key and some of the more unusual
chords we play will require special consideration.  So maybe equal temperament is the best way to
bypass the need for all this extra work; sure, it's out of tune harmonically, but how bad is it? If
we value saving time, we might need to sacrifice perfect harmony.

\subsection{You know you can't tune that instrument\ldots right?}
\label{sec:cantTune}

\begin{table}[tb]
\caption{A comparison between tunings for the twelve frequency intervals of Western music.  The
  frequency ratio for each note relative to the tonic is indicated first, followed by the interval
  size relative to the tonic in cents.  Equal temperament (ET) tuning is compared with Harmonic
  tuning (HT) often referred to as Just tuning. The final column indicates the magnitude of
  differences between the derived notes of the scales in cents. The smallest differences are
  indicated in green and the most problematic magnitudes are indicated in red.}
\label{tab:EqTempVsHarmonic}
\centering
\begin{tabular}{clcrcrrcrr}
\hline
\noalign{\smallskip}
       &              &\multicolumn{2}{l}{Equal Temperament}   
                                            &\multicolumn{2}{l}{Harmonic Tuning}
                                                                  &ET$\,-\,$HT \\
Note   &              &Frequency &Interval  &Frequency &Interval  &Difference  \\
No.    &Interval Name &Ratio     &(cents)   &Ratio     &(cents)   &(cents)     \\
\noalign{\smallskip}
\hline
\hline
\noalign{\smallskip}
 1     &Unison        &1.000  &   0.0    &$ 1/ 1$  &   0.0    & 0.0    \\
 2     &Minor 2nd     &1.059  & 100.0    &$16/15$  & 111.7    &11.7    \\
 3     &Major 2nd     &1.122  & 200.0    &$ 9/ 8$  & 203.9    & 3.9    \\
 4     &Minor 3rd     &1.189  & 300.0    &$ 6/ 5$  & 315.6    &15.6    \\
 5     &Major 3rd     &1.260  & 400.0    &$ 5/ 4$  & 386.3    &\textcolor{red}       {13.7} \\
 6     &Perfect 4th   &1.335  & 500.0    &$ 4/ 3$  & 498.0    &\textcolor{OliveGreen}{ 2.0} \\
 7     &Tritone       &1.414  & 600.0    &$ 7/ 5$  & 582.5    &17.5    \\
 8     &Perfect 5th   &1.498  & 700.0    &$ 3/ 2$  & 702.0    &\textcolor{OliveGreen}{ 2.0} \\
 9     &Minor 6th     &1.587  & 800.0    &$ 8/ 5$  & 813.7    &13.7    \\
10     &Major 6th     &1.682  & 900.0    &$ 5/ 3$  & 884.4    &15.6    \\
11     &Minor 7th     &1.782  &1000.0    &$ 7/ 4$  & 968.8    &\textcolor{red}       {31.2} \\
12     &Major 7th     &1.888  &1100.0    &$15/ 8$  &1088.3    &11.7    \\
13     &Octave        &2.000  &1200.0    &$ 2/ 1$  &1200.0    & 0.0    \\
\noalign{\smallskip}
\hline
\end{tabular}
\end{table}

Table \ref{tab:EqTempVsHarmonic} compares the Equal Temperament tuning with Harmonic tuning. The
harmonically-tuned scale presented here comprises odd harmonics up to and including the seventh
harmonic illustrated in Fig.~\ref{fig:wavemodes}. This is a common prescription called {\em 7-Limit
  Just tuning}.

This harmonic tuning is the one our ear adjusts our singing voice to, so if you want to tune your
singing voice, try perfecting Just tuning. Here we are including a Minor 7th interval with a
harmonic ratio of 7/4, which is especially important to learn if you plan to join a barbershop
quartet. The harmonic 7th chord is famous in barbershop music, and has a narrower interval from the
tonic than equal temperament by 31.2 cents.  The harmonic note is considered sweeter in quality
than equal temperament minor sevenths of a piano. 

Remarkably, our familiarity with equal temperament can fool us into thinking that it is
harmonic. Some may think that harmonically tuned chords don't sound quite right -- simply because
they are not as accustomed to them. By far the largest deviation of equal temperament from harmonic
tuning is the harmonic minor seventh, at 31.2 cents out of a possible 50 cents.  If you've always
thought 7th chords on a piano are a little unpleasant, now you know why!  They are far out of
tune. Also, by reminding people that they are perhaps familiar with equal temperament, you may
convince them that your imperfect-sounding vocals are in perfect harmony.

The real problem exposed in Table \ref{tab:EqTempVsHarmonic} is the major 3rd. It is very prominent
in modern music, and yet we see here that it is {\em far} from the near perfection of the perfect
4th and 5th intervals.  As the fifth harmonic in Fig.~\ref{fig:wavemodes}, the major 3rd is a
relatively loud harmonic. Imperfections of the scale tuning for the major 3rd interval will clash
with the harmonic overtones of the tonic note of the scale. To see the problem intuitively, we can
visualise the difference between the smooth harmonies of Just tuning and the quick compromise of
equal temperament, using plots called Lissajous curves. Check out the plots in
Ref.~\cite{SierraForthcoming-SIERIL}, and the YouTube video in Ref.~\cite{Entonal_2022}, and even
make your own using Ref.~\cite{monman53}. To survive this problem, we will need some creative
adjustments.

\subsection*{So, what is the solution?}

If you are recording, there is a very nice solution.  With modern software, you can record in equal
temperament and then let the software make the small adjustments to harmonic tuning.  As an example
of this, see mm:ss=04:40 of the YouTube video of Ref.~\cite{MelodyneTuning}.

If you are playing live, and your instrument can be readily tuned, it is best to undo some of the
compromises made in equal temperament and get a little closer to harmonic tuning, as much as the
song allows.  Table \ref{tab:EqTempVsHarmonic} provides a guide to the most problematic notes.  

You'll need a fresh guitar for every song with a slightly different tuning, but we do certify this as
a completely valid reason for buying another guitar.  Official certificates stating this need are
available at URL of Ref.~\cite{Certificate}.

\begin{figure}[tb]
  \begin{center}
  \includegraphics[width=0.32\textwidth]{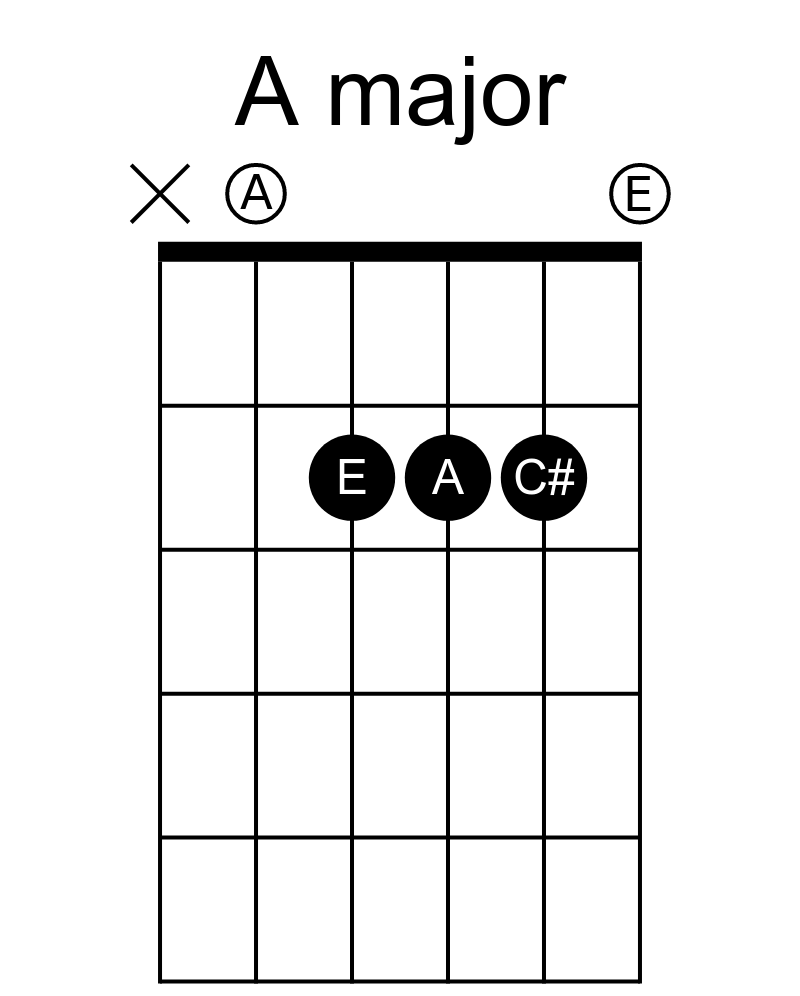}
  \qquad
  \includegraphics[width=0.32\textwidth]{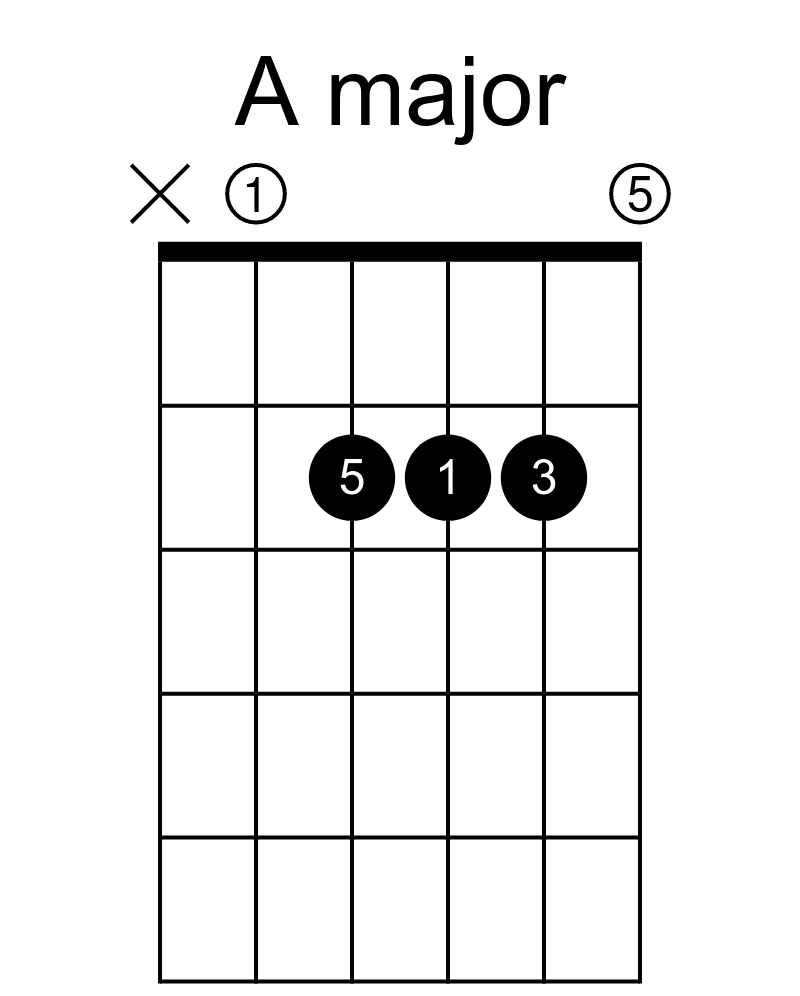}
  \end{center}	
  \caption{Chord chart for an A major chord on a 6-string guitar in standard tuning.  Note names
    (left) and note intervals from the A tonic (right) are indicated. {\large $\times$} denotes the
    muting of the sixth string.}
  \label{fig:Achord}
\end{figure}

So, how do you handle the tuning?  Consider, for example, the open A chord with the 1st and 5th
strings open and the 2nd, 3rd, and 4th strings fretted at the second fret as illustrated in
Fig.~\ref{fig:Achord}.  Here the second string is playing a C-sharp, the major 3rd in the A
scale. It sounds awful because the equal-tempered 3rd is sharp by a lot, at 13.7 cents as indicated
in Table \ref{tab:EqTempVsHarmonic}.  So go ahead, use that guitar tuner as a first step, but don't
stop there.  Finish the job.  The pros do.  Use your ear and flatten the 2nd (B) string until it
sounds okay, but not too far, or every other chord will then be out of tune\ldots.  You see, you
really can't tune that instrument. For a deeper look at the problem on a piano, take a look at
Ref.~\cite{minutephysics_2015}.

The imperfections of equal temperament become even more important if we decide to amplify our
sounds with a little distortion. We will see how the harmonic content of our notes determines
whether distortion sounds good or nasty. The pros certainly know the importance of using harmonic
tuning, although they may not call it that. For example, Eddie Van Halen would tune his guitar by
ear to what sounds good, and then ask the rest of the band to tune to him. Apparently, most Van
Halen songs are not recorded with reference to A 440. 

So, if we persevere, we can select a fairly satisfying 12-tone scale and start to make
music. However, when it comes to sending signals into a guitar amplifier, the result may sound
confusing. This is because amplifiers add extra frequencies with an intricate dependence on the
input frequencies. If the input signal is tuned harmonically, the overtones align well, but slight
differences will compound into a much more chaotic amplifier output. Not only is the volume
stronger, but there are now also several types of distortion frequency patterns.

\section{Loud amplifiers produce strange sounds}
\label{sec:distortion}

Our brains seek out those joyous {\em frequency ratios of small integers} in whatever music we hear
or play, even when listening to many chords at once. But as when we listen to the timbres of
different musical instruments, simplicity alone cannot explain the beauty of music - sometimes our
chord combinations are by design less like {\em chord sandwiches} than they are like shockingly
large burgers, filled with layer after layer of sauce-drenched ingredients. For example, so many
frequencies distort the sound when we listen to loud music through valve amplifiers.

\subsection{Nonlinear Amplification}

A valve (or tube) amplifier driven loud gently rounds off the peaks of the input wave form.
Instead of faithfully reproducing the input wave form, the valves saturate and are unable to reach
the original peaks. This is the origin of the guitar-amplifier distortion celebrated by enthusiasts
of rock, pop, electronica, and almost any other genre nowadays. A distorted amplifier will leave us
with a completely different sound from the one we started with -- which may or may not be
desirable.

But how does this come about - what is happening to the music? Is distortion simply changing the
amplitudes of the overtones for the notes, or is it more complicated? As you might guess, it is
wonderfully complicated!

We want to figure out which frequencies are present in the voltage that comes out of the amplifier,
so let us give that a name: $V_{out}$, a function of the input voltage $V_{in}$. A linear
amplification function gives back what we put in, but louder, for example,
\begin{equation}
V_{\rm out} = 10 \, V_{\rm in} \, ,
\end{equation}
illustrating a {\em perfect} amplifier. Here, the leading factor of 10 provides an order of magnitude
amplification of $V_{\rm in}$. It increases the amplitude of all input voltages by the same
relative amount.

\begin{figure}[tb]
	\includegraphics[width=0.99\textwidth]{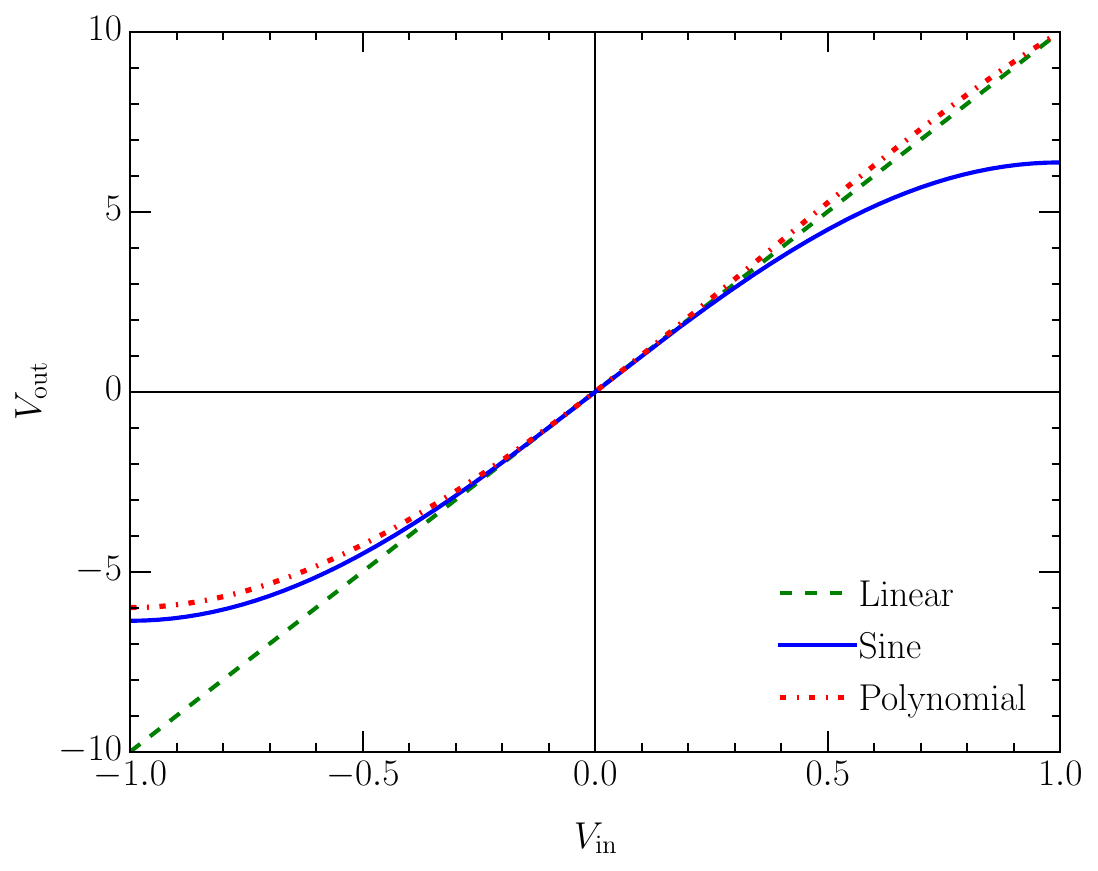}
	\caption{Three amplifier functions producing different output, $V_{\mathrm out}$, as a
          function of the input voltage, $V_{\mathrm in}$.  The linear amplification function
          illustrates a perfect amplifier, reproducing the input signal faithfully.  The
          nonlinear functions are representative of the amplification curve for a valve amplifier,
          modelled on a sine function or polynomial as described in the text.  For small signals, the
          amplifier is almost perfect.  However, when the amplifier is hit with large input
          signals, by digging into the guitar strings for example, the amplifier saturates and
          rounds off the peaks of the input wave form.}
	\label{fig:linnonlinamp}
\end{figure}

Nonlinear amplification has a more complicated dependence on the input voltage, such as the two
curved amplifier functions in Figure~\ref{fig:linnonlinamp}. The nonlinear curves are
representative of a valve amplifier.  The curves are approximately linear for small input signals,
where the amplifier is almost perfect.  But when the amplifier is hit with large input signals, the
input voltage is amplified to an output voltage that is different in magnitude than that required
for perfect amplification. For the sine function for example
\begin{equation}
V_{\rm out} = 10 \, \frac{2}{\pi}\, \sin \left( V_{\rm in} \, \frac{\pi}{2} \right) \, ,
\label{eq:sine}
\end{equation}
the peaks in an input wave form will be flattened, introducing distortion.  Here, the leading factor
of 10 provides an order of magnitude amplification, while the next factor ensures the
slope of the amplification function is 1 at $V_{\rm in} = 0$.  This ensures near perfect
amplification for small input voltages.

The third amplification function in Figure~\ref{fig:linnonlinamp} is a polynomial with
\begin{equation}
V_{\rm out} = 10 \, \left ( V_{\rm in} + a\, V_{\rm in}^2 - a\, V_{\rm in}^3 \right ) \, ,
\label{eq:poly}
\end{equation}
where the matching amplitudes of the nonlinear terms, $a$, ensure that $V_{\rm out} = 10$ at
$V_{\rm in} = 1$. Demanding a zero slope for the amplification function at $V_{\rm in} = -1$ sets
$a=1/5$ which we use herein. The $V_{\rm in}^2$ term is symmetric, which breaks the perfect
anti-symmetry of the linear and cubic terms, giving us different behaviour at positive and negative
$V_{\rm in}$. This will be of utility as we learn how nonlinear amplification generates new notes
not played by the musician.
%

Whenever the amplifier function is curved, the wave is squashed. The curve sets a lower output
voltage, which prevents the wave from reaching its fully amplified height, like adding less
fertiliser to some parts of a garden bed. Even though the entire garden might be growing the same
type of plant, the plants with less fertiliser might have stunted growth.

The amplifier functions appear to make sense in terms of voltages, but to understand their effect
on music notes, in terms of patterns in frequencies and harmonies, we need to check what happens to
the input wave forms. Sound waves are converted into electrical signals by microphones, which is
when they obtain a voltage depending on several factors, including the loudness of the
sound. Similarly, guitar pickups for example convert the motion of a vibrating guitar string into a
voltage. So, notes (and their harmonics) that are played at louder volume on an instrument will
generate higher voltages in an amplifier.

Modern guitar amplifiers have several gain stages to allow the musician to decide just how hot the
signal is as it goes into he next valve stage. By dialling in just the right amount of gain on the
amplifier, one can play clean sounds with little distortion when the notes are played gently. But,
when the musician digs in and plays hard, the input signal enters the nonlinear part of the
amplification and musical wonders unfold!

\subsection{Wave Form Modification}

\begin{figure}[tb]
	\includegraphics[width=0.32\textwidth]{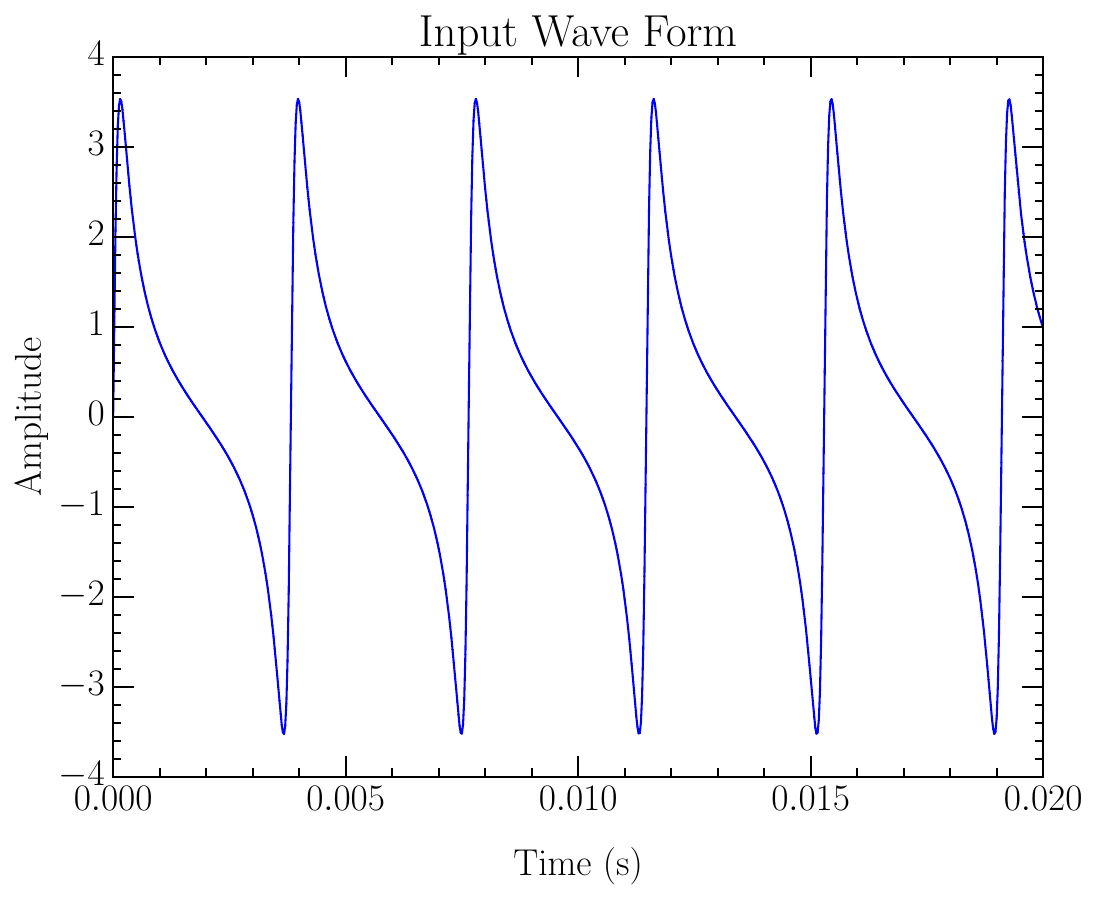}
	\includegraphics[width=0.32\textwidth]{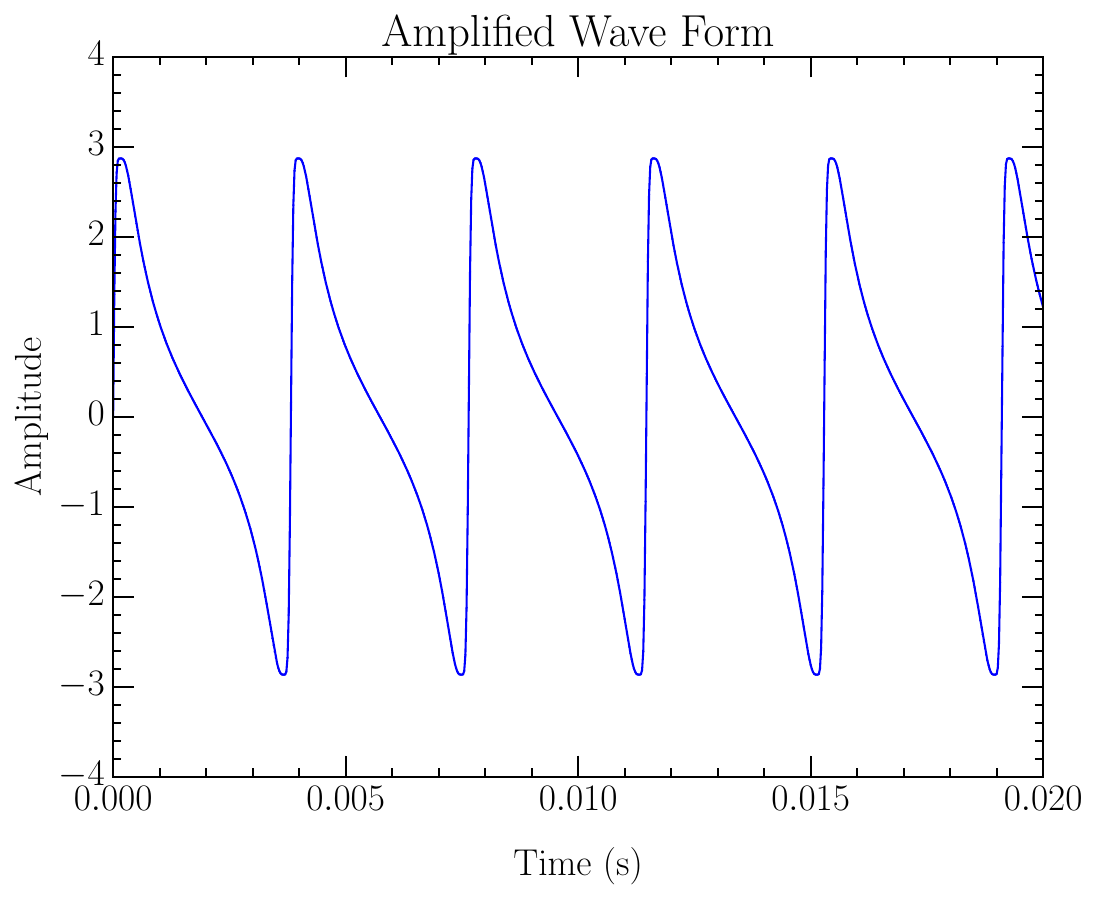}
	\includegraphics[width=0.32\textwidth]{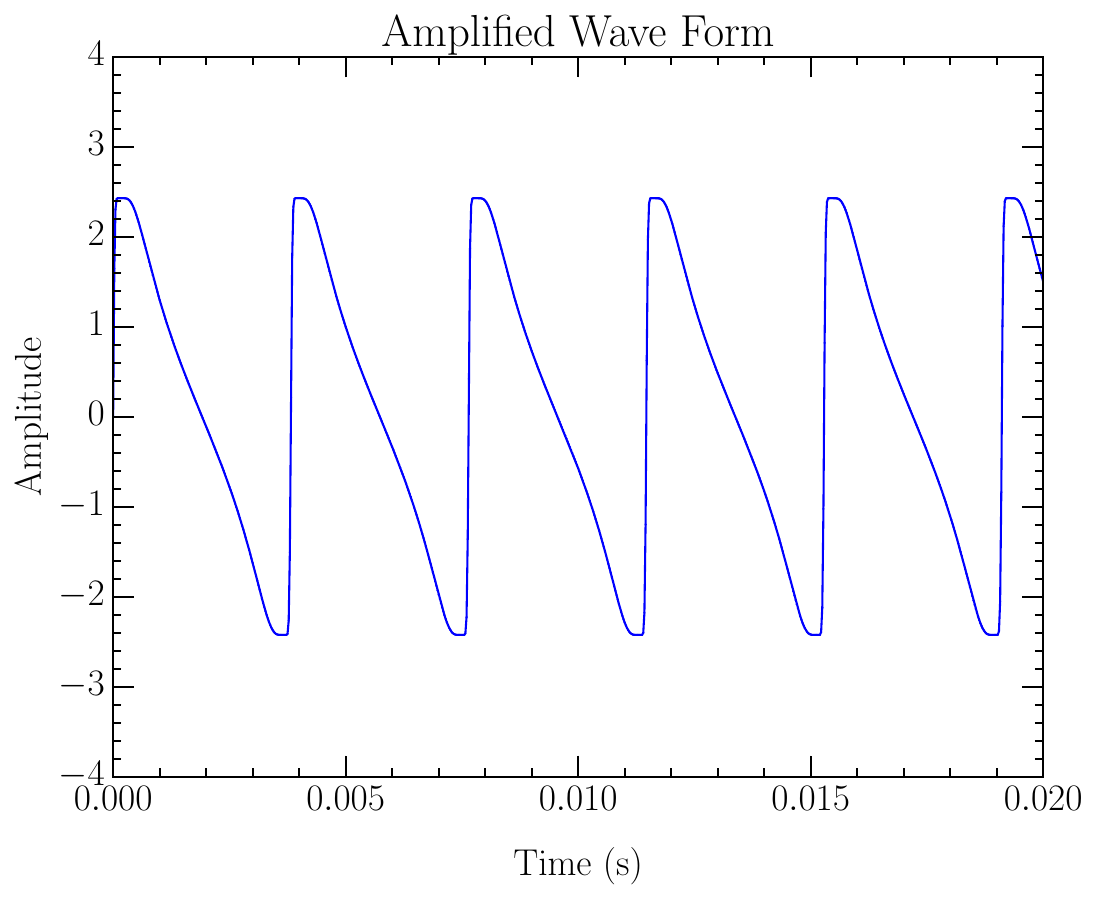}
	\caption{Wave forms for a C4 note as depicted in the power spectrum of
          Fig.~\ref{fig:firstpowerspectrum} illustrated over a duration of 2 one-hundredths of a
          second.  The input wave form (left) is amplified by an amplifier modelled by a sine
          function (middle), and then amplified again to compound the nonlinear effects as in an
          amplifier with multiple gain stages (right).  The RMS amplitude is held fixed to
          illustrate changes in the wave form due to nonlinear amplification.}
	\label{fig:waveformsine}
\end{figure}

Figure \ref{fig:waveformsine} illustrates the wave form for the C4 note (an octave below middle C)
as described in Fig.~\ref{fig:firstpowerspectrum} in the left-hand plot.  Note how the harmonic
series combines to create a wave form quite different from a familiar sine wave. The middle
plot illustrates how the wave form has changed after a pass through the amplifier modelled by a
simple sine function.

Just as real amplifiers typically have more than a single gain stage, the amplified signal can be
put through the amplifier again to compound the effects.  For illustrative purposes, the second
gain stage uses the amplified signal from the first stage, normalised to maintain $-1 \le V_{\rm
  in} \le 1$.  In the plot of the amplified signal, the root-mean-square (RMS) amplitude of the
input wave form is maintained, to illustrate differences in the shape of the wave form as opposed
to the usual amplitude gain achieved by a real amplifier.

\begin{figure}[tb]
	\includegraphics[width=0.32\textwidth]{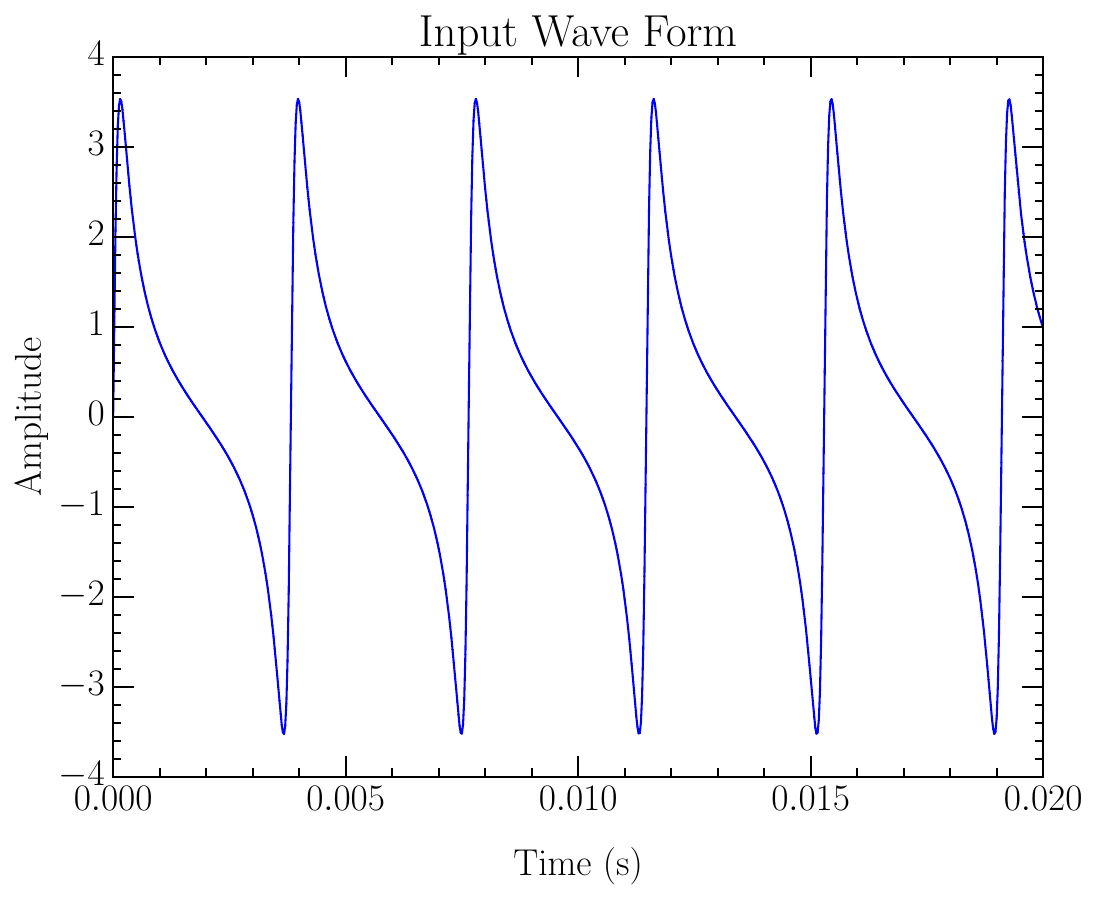}
	\includegraphics[width=0.32\textwidth]{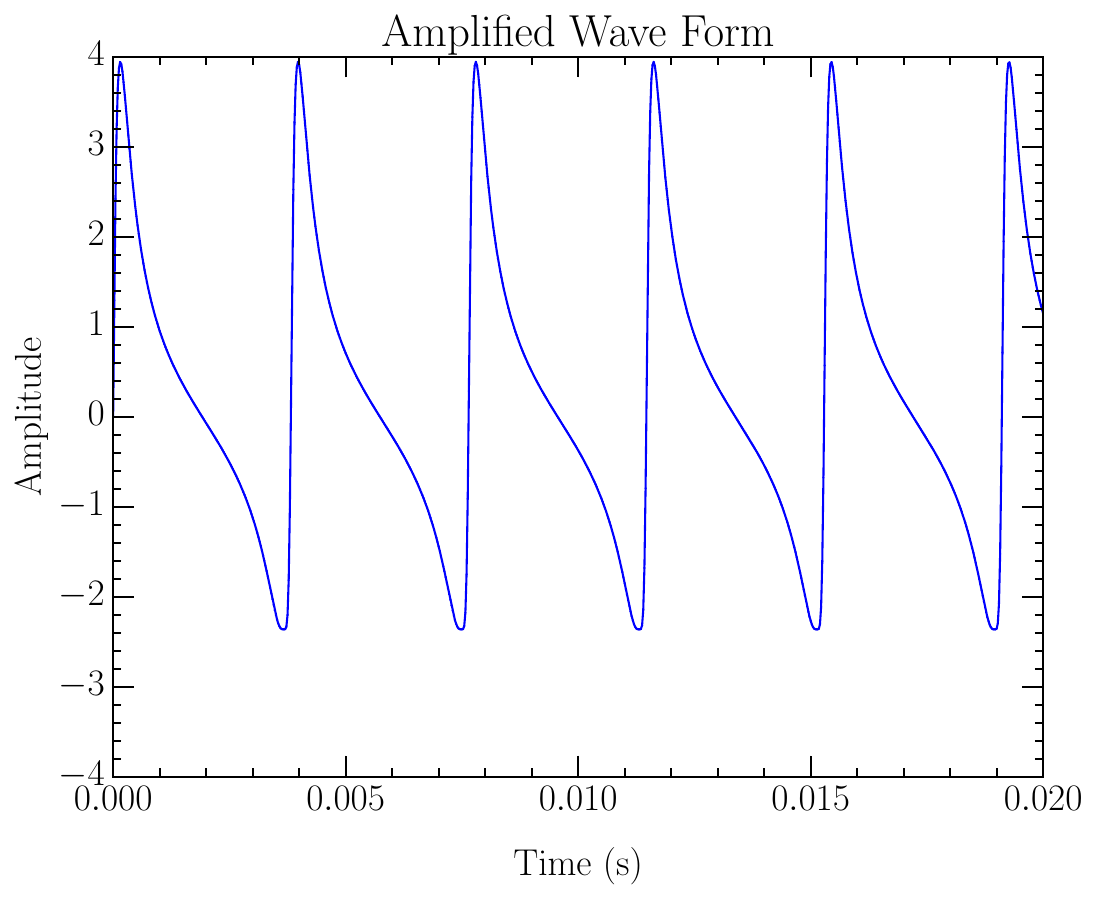}
	\includegraphics[width=0.32\textwidth]{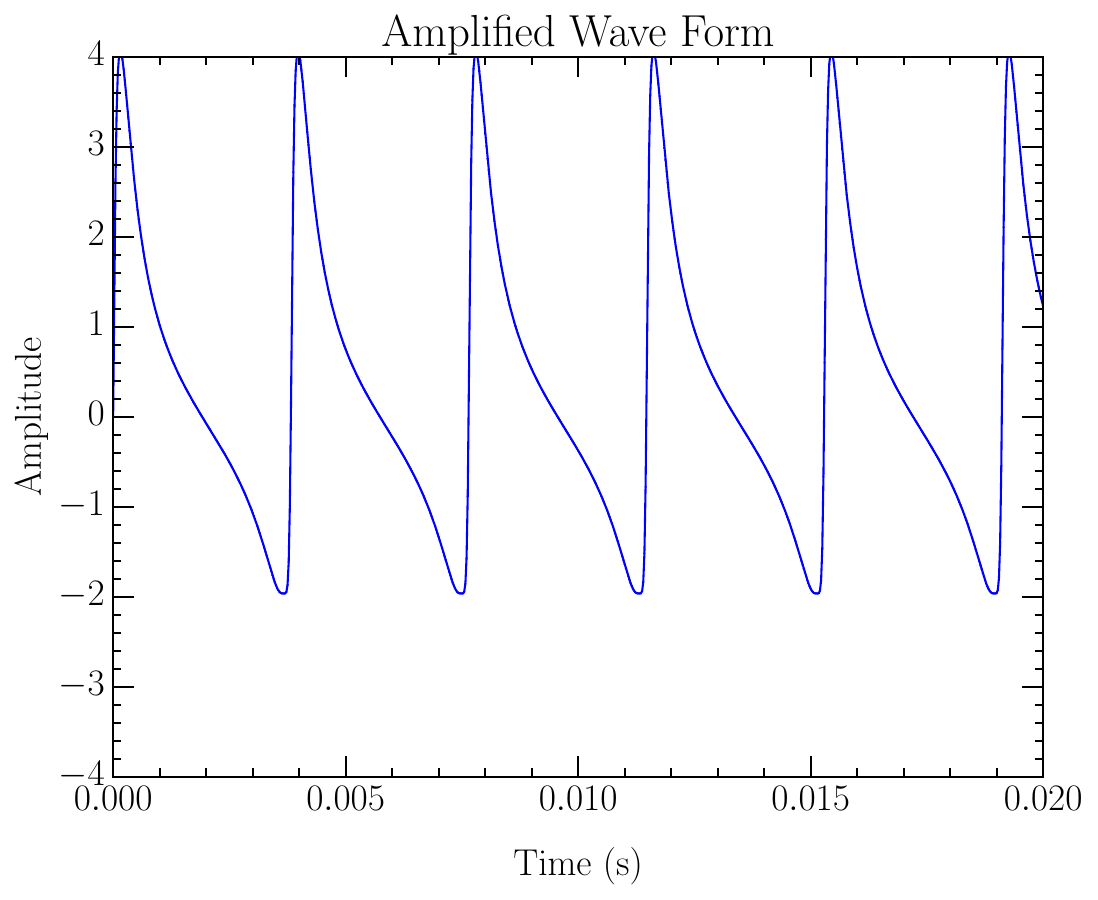}
	\caption{As in Fig.~\ref{fig:waveformsine} but for an amplifier modelled by the polynomial
          of Eq.~(\ref{eq:poly}).  }
	\label{fig:waveformpoly}
\end{figure}

Figure \ref{fig:waveformpoly} illustrates the wave form for the same C4 note 
amplified this time by the amplifier modelled by the polynomial of Eq.~(\ref{eq:poly}).

\subsection{Power Spectrum Modifications}

\subsubsection{Exploring the frequencies present in the wave form}

To understand what this distortion of the input signal is doing musically, we need to analyse the
frequencies in the amplifier's output signal and compare them to the frequencies in the input
signal. Similar to light, musical frequencies form a spectrum, and human hearing is sensitive from
about 20 Hz to 20,000 Hz \cite{Purves2001Neuroscience}.  Our aim is to determine how much of each
audible frequency is present in the input and output signals. This allows us to create a power
spectrum\footnote{In simple mathematical terms, the power spectrum is the absolute-square Fourier
transform of the input wave form. For an entertaining look at the power of a Fourier transform,
check out the YouTube video in Ref.~\cite{Veritasium_2022}.}, as we did in
Fig.~\ref{fig:firstpowerspectrum} where we explored the harmonic power spectrum for a C4 note on a
piano string. For a physics-of-music-style description of Fourier transforms for signal processing,
try Chapter 11 of this book \cite{dspguide}.

\subsubsection{Mathematical details}

For completeness we describe the mathematical details of the calculations presented here.  If this
isn't of interest to you, feel free to skip to the next section.

In constructing the wave form presented in the left-hand plot of Fig.~\ref{fig:waveformsine} we
commence by selecting the equal-tempered fundamental frequency of the piano C4 note an octave below
middle C.  In standard tuning with the A4 fundamental frequency at $f_{\rm A4} = 440$ Hz, the A3
note has a frequency of $f_{\rm A3} = 220$ Hz.  As C4 is 3 semitones above A4, the frequency is
\begin{equation} 
f_{\rm C4} =  f_{\rm A3} \, 2^{3/12} = 261.63\ \mbox{Hz} \, .
\end{equation}
We then consider $n_h=30$ harmonics at integer multiples of the fundamental frequency.  
\begin{equation} 
f_i = f_1 \times i \, .
\end{equation}
As we will implement an exponential fall off in the strength of the overtone series, this is
generous.

The power of a sine wave with amplitude $A$ is $A^2/2$.  Thus to set the power of the fundamental
at 1, we set the amplitude of the fundamental tone $A_1 = \sqrt{2}$. The harmonics are given
an exponentially decaying amplitude
\begin{equation} 
A_i = A_1\, \exp\left( -\sigma \, \frac{f_i-f_1}{f_1} \right) \, ,
\end{equation}
with $\sigma$ determined to suppress the amplitude of the 10th harmonic to $s = 1/10$th of the first
harmonic 
\begin{equation} 
\sigma = -\frac{f_1}{f_{10}-f_1}\, \ln( s ) \, ,
\end{equation}
where reference to $f_{10}$ corresponds to our selection of the 10th harmonic for consideration.

With the frequencies and corresponding amplitudes selected, the wave form is constructed.  We
create the wave form for a long $T = 4$ second duration or more. This will provide excellent
resolution in the discrete power spectrum analysis with a fine spacing between frequencies and
little signal leakage associated with the finite size of the time interval.

To address frequencies up to 20,000 Hz, the upper limit of human hearing, we set the Nyquist
critical frequency $f_c = 20,000$ Hz.  As the sampling is done twice per period or cycle the
sampling frequency, $f_s$, is twice the critical frequency
\begin{equation} 
f_s = 2\, f_c = \frac{1}{\delta t} \, , \quad\mbox{such that}\quad 
\delta t = \frac{1}{2\, f_c} \, ,
\end{equation}
where $\delta t$ is the time interval between samples, $25\, \mu s$ in our case. Thus the number of
samples calculated is $n_s = T/\delta t = 160,000$ samples for $T=4\,$s.  However, the fast Fourier
transform algorithm we use herein requires the number of samples to be a power of 2.  As a result,
the number of samples considered is
\begin{equation} 
N_s = 2^{\left\lceil {\ln n_s}/{\ln 2} \right\rceil} = 2^{18} = 262,144\ \mbox{samples} \, ,
\end{equation}
a duration of $6.5536\,$s. The wave amplitude at sample time $t_i = \delta t \, (i-1)$ is then given
by
\begin{equation} 
V(t_i) = \sum_{j=1}^{n_h} \, A_j \, \sin \left ( 2\pi\, f_j \, t_i \right ) \, .
\label{eq:wave}
\end{equation}
Several notes can be combined by performing the calculation of Eq.~(\ref{eq:wave}) for each note
and accumulating the results in $V(t_i)$.  

The power spectrum is calculated via a fast Fourier transform.  For $N_s$ samples, there are $N_s/2
+ 1$ positive frequencies (including zero) with a frequency spacing of
\begin{equation} 
\delta f = \frac{1}{ N_s\, \delta t } = \frac{2\, f_c}{N_s} \simeq 0.153\,\mbox{Hz.}  
\label{eq:freqSpacing}
\end{equation}
Thus, the $k$'th frequency is $f_k = (k-1) \, \delta f$ with $k = 1,\, 2,\, 3,\ldots N_s/2 + 1$,

Denoting the complex Fourier coefficient for the $k$'th frequency as $F_k$,
the power spectrum $P_k$ is
\begin{equation} 
P_k = \frac{\vert F_k \vert^2}{N_s^2} \, .
\label{eq:power}
\end{equation}

As the frequencies available in the discrete transform do not align perfectly with the input
frequencies considered in creating the wave form, spectral strength is often divided between
adjacent frequencies in the transform.  To ensure the power spectrum plots reflect the full
strength associated with a given input frequency, we sum the power at each discrete frequency with
its adjacent frequencies over a range of 0.6 Hz, a range large enough to capture the full strength
of the input wave form.  For example, this ensures a sine wave with amplitude $\sqrt{2}$ has a
power-spectrum plot with strength 1 at the input frequency.

Amplification via the curves of Fig.~\ref{fig:linnonlinamp} proceeds by finding the maximum
absolute-value amplitude of the input wave form $V_{\rm max} = \max \vert V(t_i) \vert$,
normalising the input wave $V_{\rm in}(t_i) = V(t_i) / V_{\rm max}$ such that $-1 \le V_{\rm
  in}(t_i) \le 1$ and then applying the amplification function.  For the simple sine function
\begin{equation} 
V_{\rm out}(t_i) = \sin\left ( V_{\rm in}(t_i) \, \frac{\pi}{2} \right ) \, .
\end{equation}
For the polynomial
\begin{equation} 
V_{\rm out}(t_i) = V_{\rm in}(t_i) + a\, V_{\rm in}^2(t_i) - a\, V_{\rm in}^3(t_i) \, .
\end{equation}
After the amplification, the wave form is normalised to preserve the RMS strength of the input wave
\begin{equation} 
V_{\rm RMS} = \sqrt{ \frac{1}{N_s} \, \sum_{i=1}^{N_s} V^2(t_i) } \, .
\end{equation}

\subsubsection{Power spectrum of a distorted amplifier: single note}

\begin{figure}[tb]
	\includegraphics[width=0.32\textwidth]{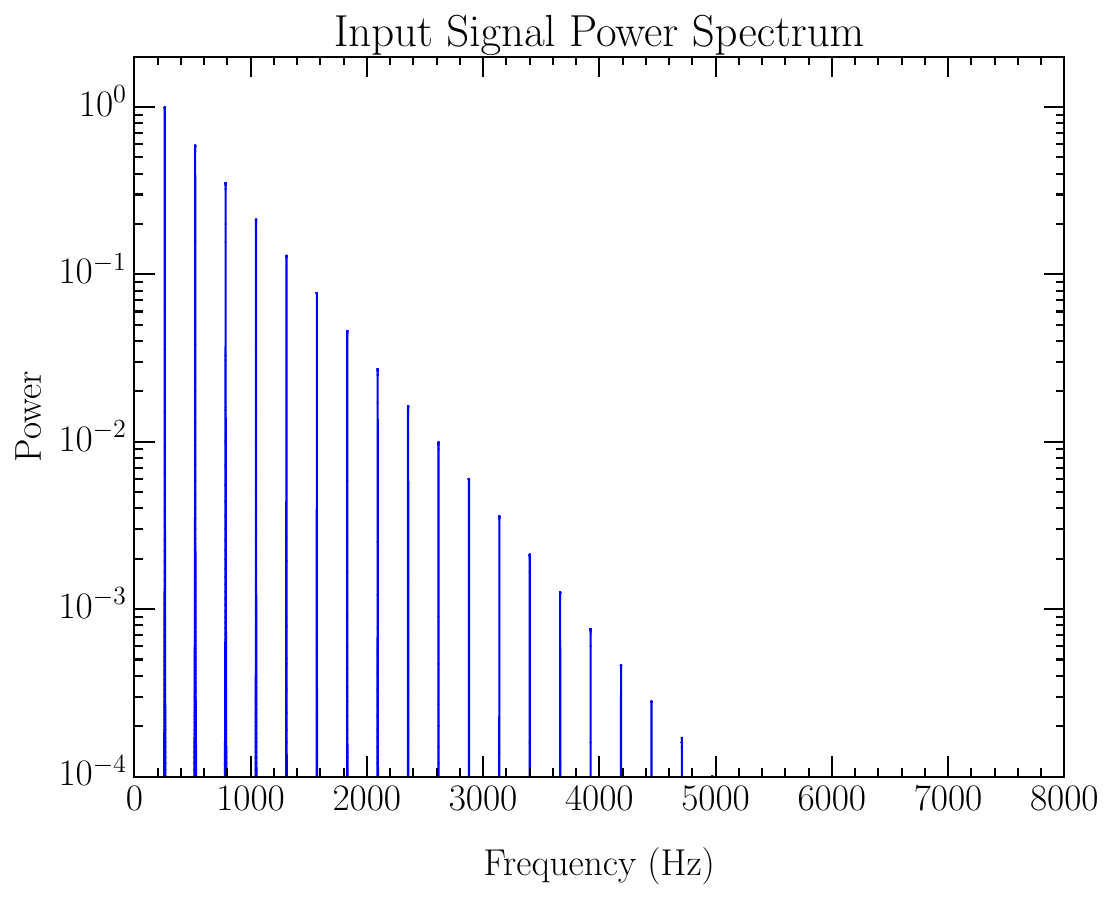}
	\includegraphics[width=0.32\textwidth]{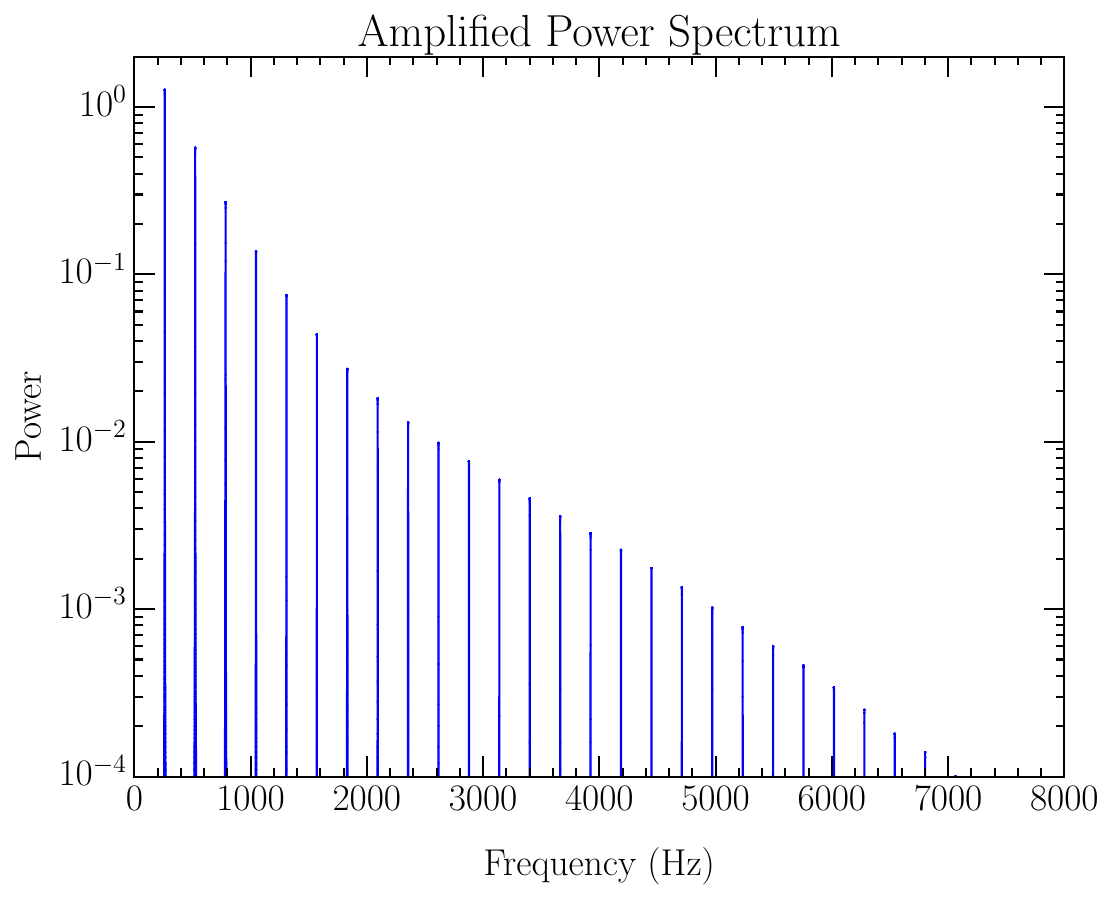}
	\includegraphics[width=0.32\textwidth]{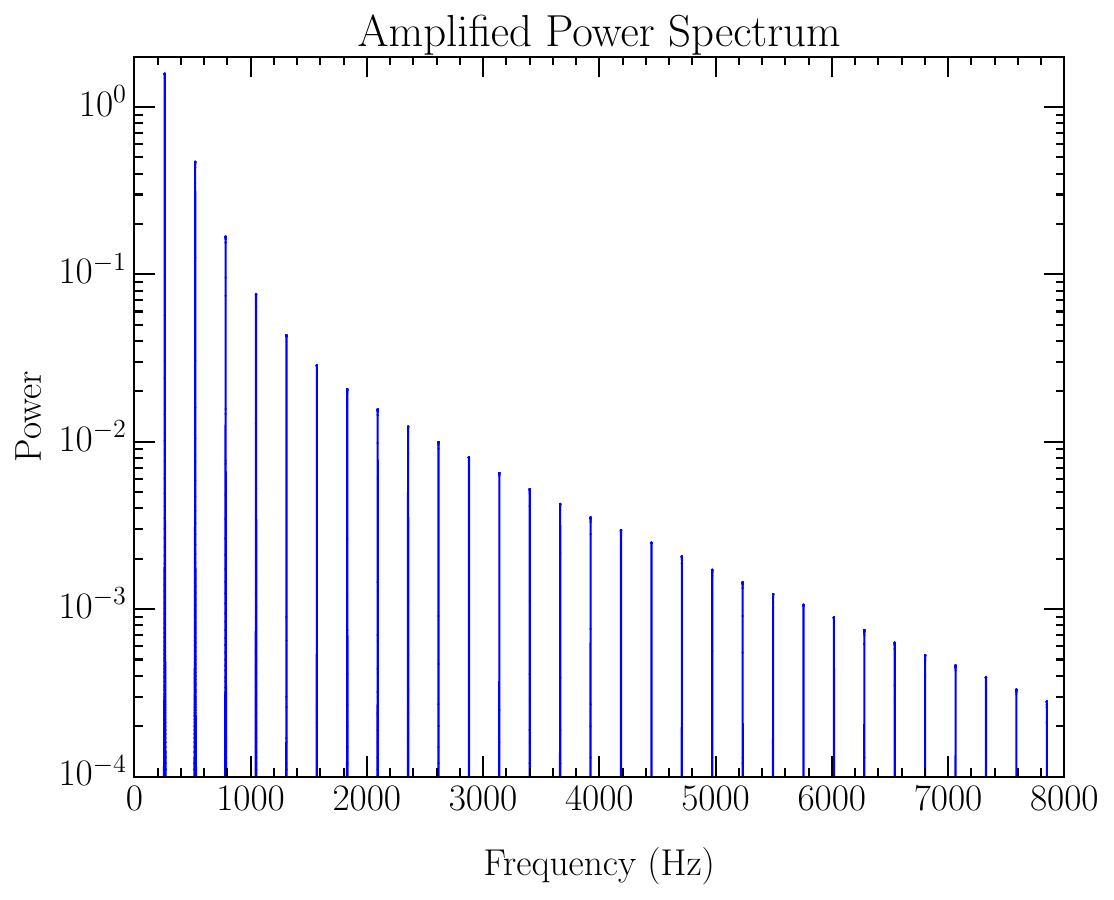}
        \caption{Power spectra for the input wave form (left), output wave form after one gain
          stage (middle) and output wave form after a second gain stage for an amplifier modelled
          by the sine function. The growth of significant power in the higher frequencies will make
          the single note sound brighter after amplification.}
  \label{fig:1note}
\end{figure}

\begin{figure}[tb]
	\includegraphics[width=0.32\textwidth]{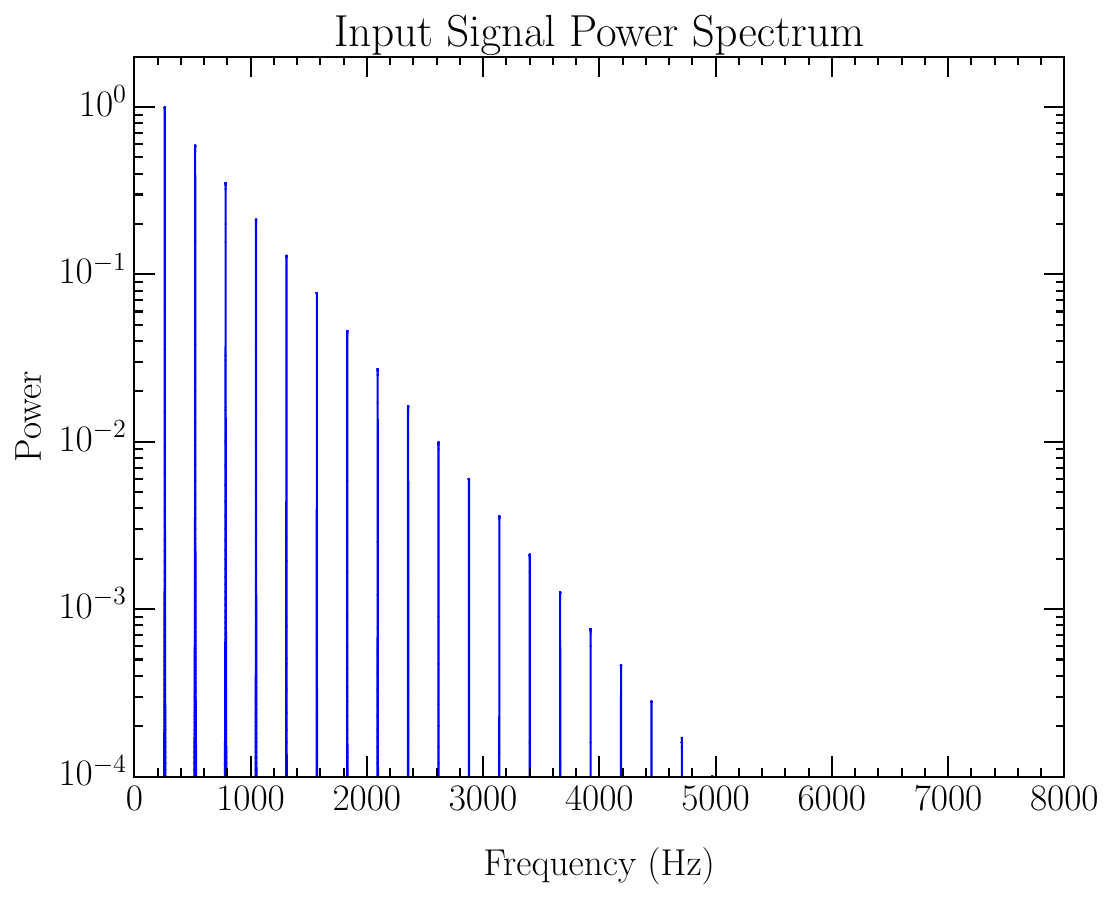}
	\includegraphics[width=0.32\textwidth]{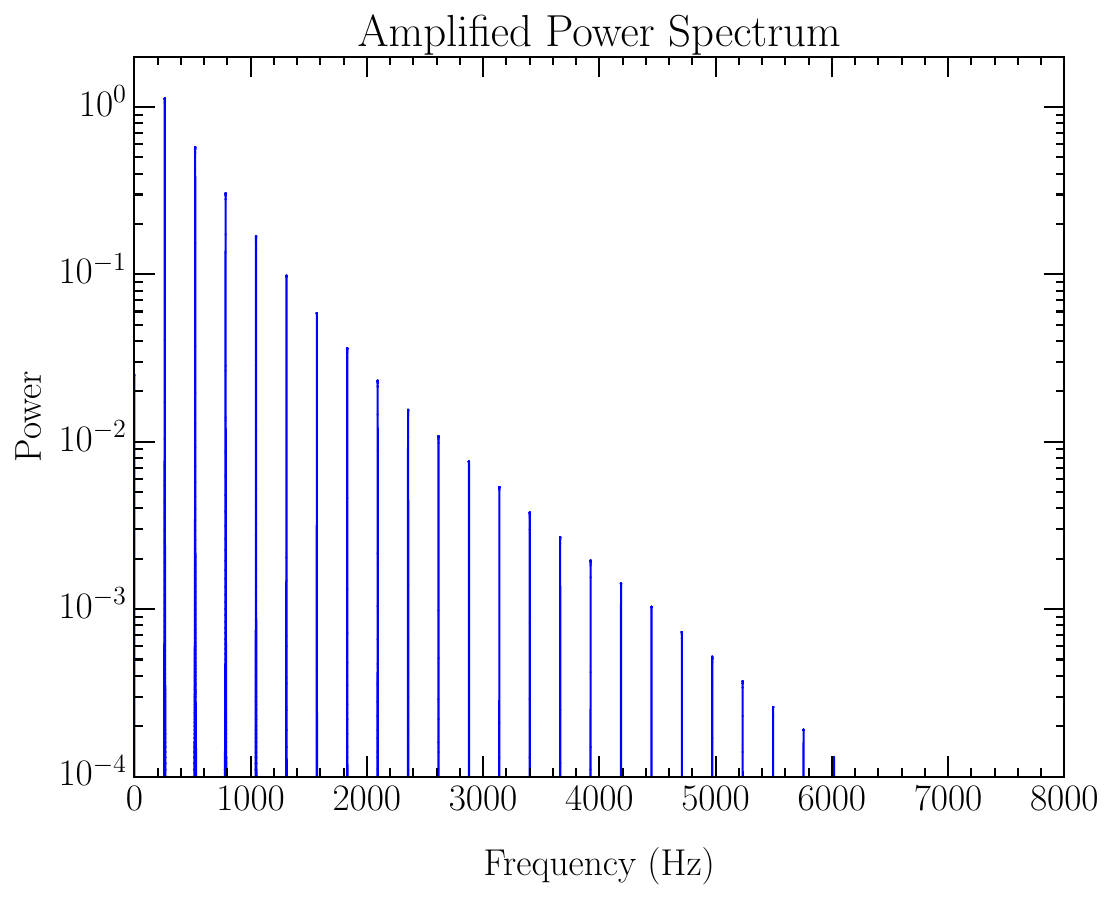}
	\includegraphics[width=0.32\textwidth]{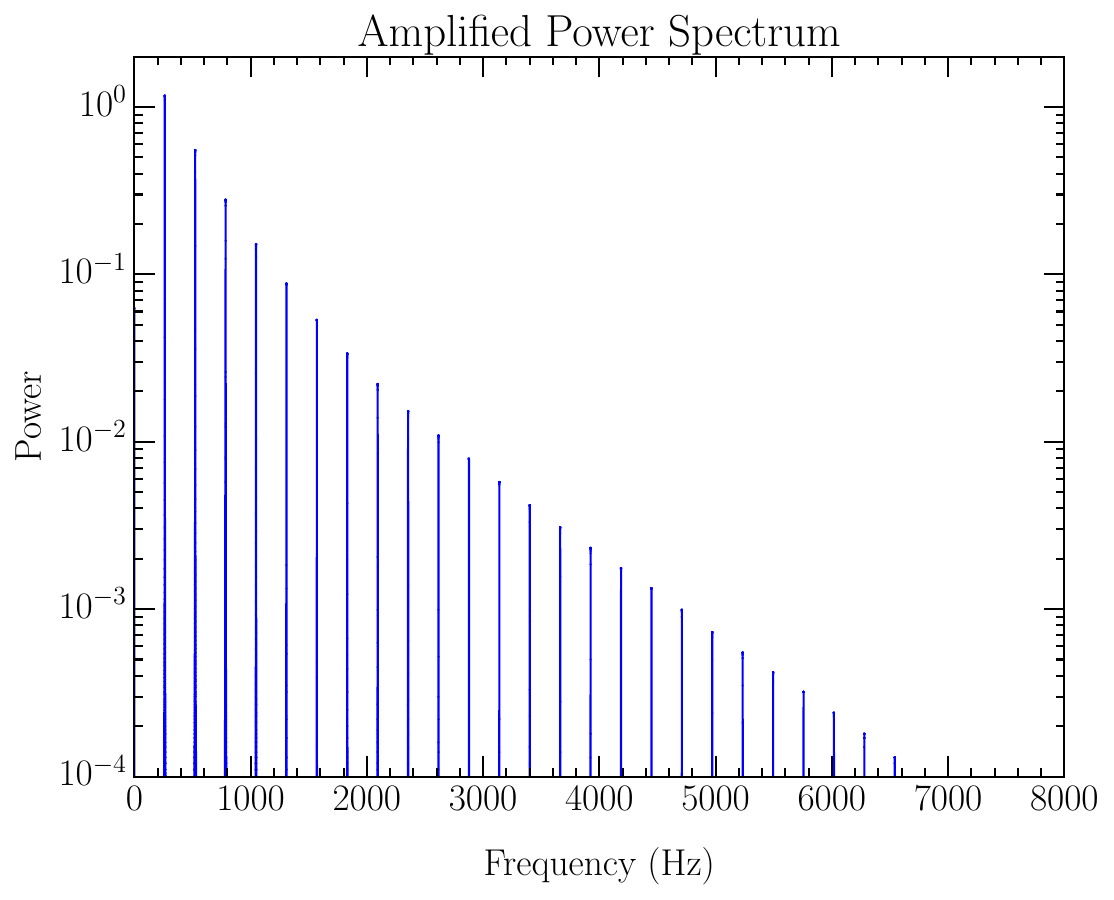}
	\caption{As in Fig.~\ref{fig:1note} but for an amplifier modelled by the polynomial
          of Eq.~(\ref{eq:poly}).  }
	\label{fig:1notepoly}
\end{figure}

Because the results are wonderfully complicated, let us start in a simple manner by considering the
amplification of a single note composed of the fundamental tone plus the overtones with an
exponentially decaying amplitude similar to that in Fig.~\ref{fig:firstpowerspectrum}.  Recall that the
overtones create the timbre. 

We continue with the C4 note an octave below middle C.  Its original power spectrum is illustrated
in the left plot of Fig.~\ref{fig:1note}.  This time the spectra are plotted on a logarithmic
scale. The exponential fall off of the overtone amplitudes displays linearly on the logarithmic
scale of the $y$-axis.  Our ears hear music loudness in this manner, such that it is an intuitive
way to illustrate the results of amplification.

After amplifying this wave form illustrated in the left-hand plot of Fig.~\ref{fig:waveformsine},
through the sine amplification curve of Fig.~\ref{fig:linnonlinamp} and Eq.~(\ref{eq:sine}), we
obtain the power spectrum for $V_{\mathrm out}$. Figure~\ref{fig:1note} provides a side-by-side
comparison of the input (left), first gain-stage power spectrum (middle) and second gain-stage power
spectrum (right). We see that for a single note with a perfect harmonic overtone series, the effect
is to enhance the power of the higher overtones relative to the original power
distribution. The amplified sound still sounds like a single note but becomes brighter, developing
a bright timbre.

Figure~\ref{fig:1notepoly} provides the same comparison for the amplification modelled by the
polynomial curve of Fig.~\ref{fig:linnonlinamp} provided in Eq.~(\ref{eq:poly}).

\subsubsection{New notes through distortion: Amplification of two notes}

\begin{figure}[tb]
  \includegraphics[width=0.49\textwidth]{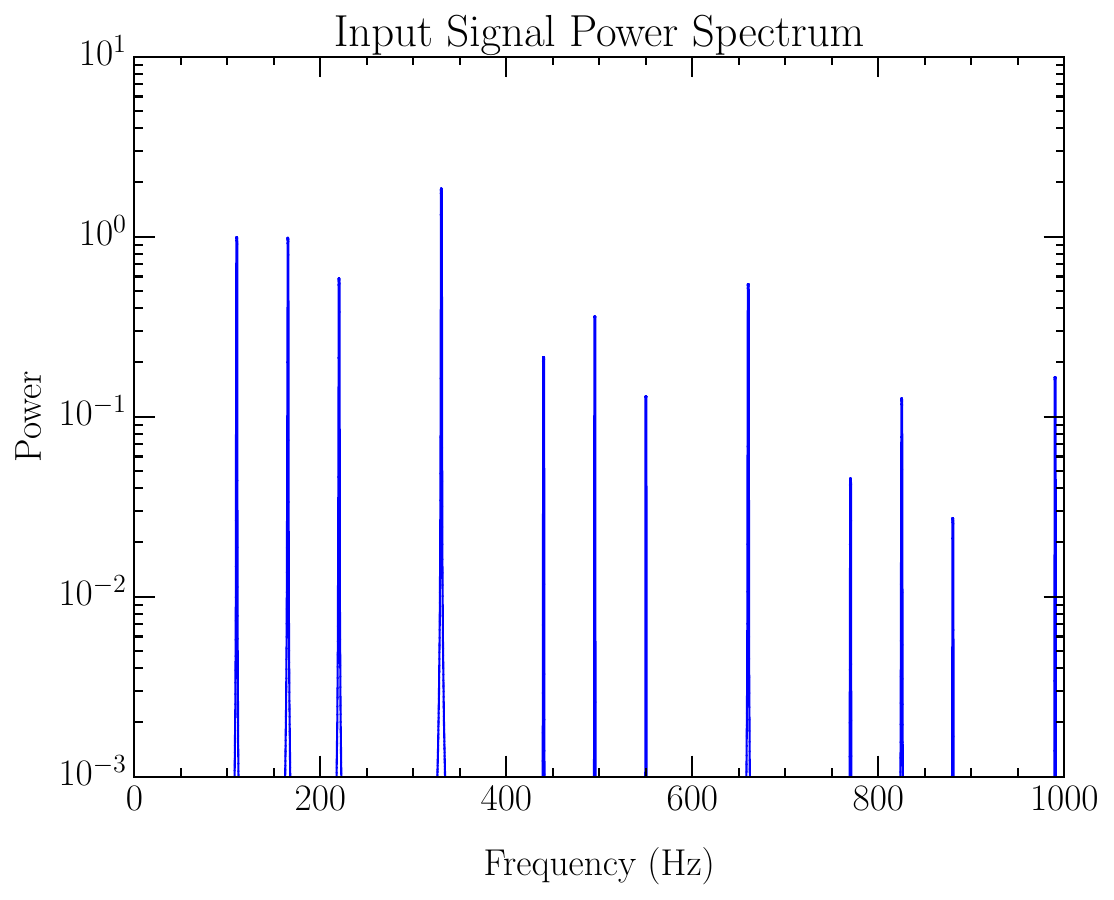}
  \includegraphics[width=0.49\textwidth]{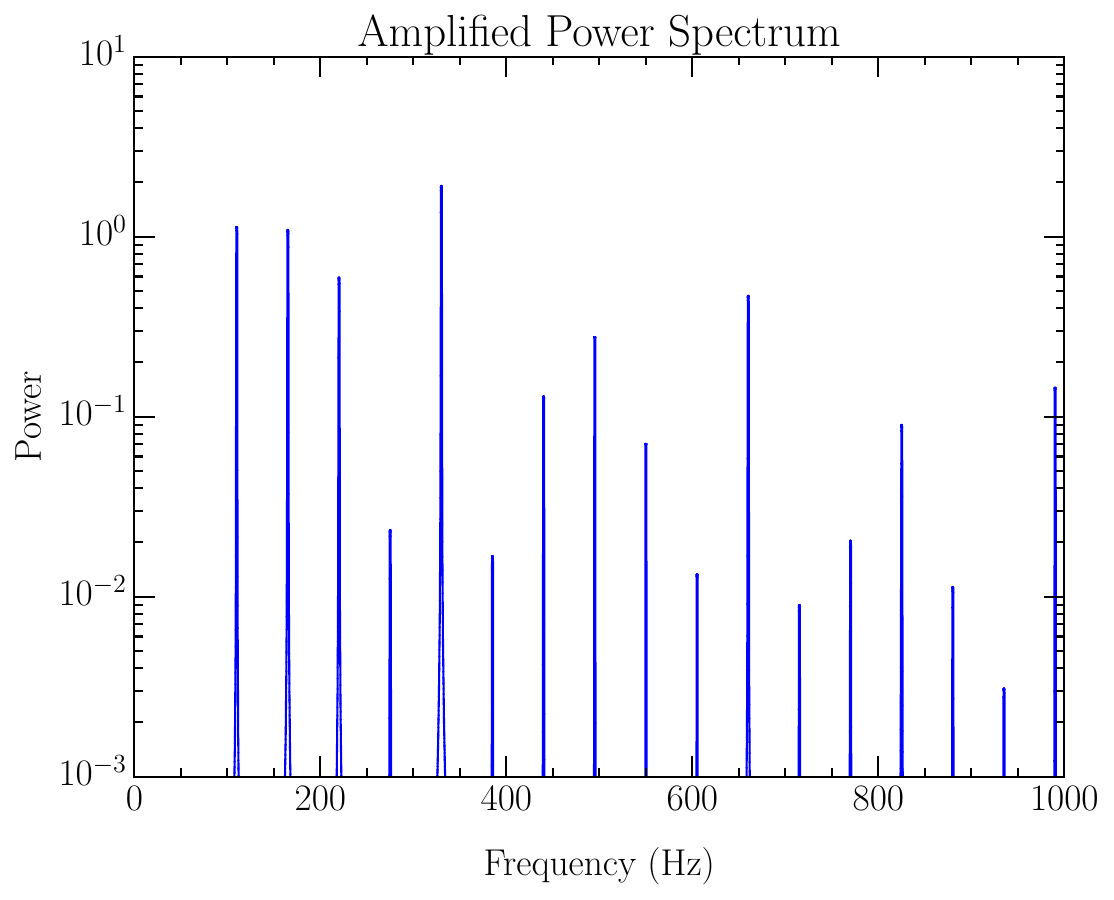}
  \caption{Power spectra for input (left) and output (right) wave forms after nonlinear sine
    amplification.  This time two notes are played with equal amplitudes, the A 110 Hz open 5th
    string and the E 165 Hz played at the 2nd fret on the 4th string.  Note how distortion via
    nonlinear amplification has generated new tones in the harmonic series.  }
  \label{fig:2notesSine}
\end{figure}

Now let us amplify two notes. Let's consider the power-chord foundation of rock, the A 110 Hz open
5th string and the E played at the 2nd fret on the 4th string.  Following Eddy Van Halen, we'll
tune the fourth string listening to the A-E combination to get it to a true perfect fifth, {\it
  i.e.} the ratio of frequencies will be 3/2 exactly such that the E note frequency is $110 \times
3/2 = 165$ Hz. Start with a guitar tuner and then use your ear to make it sound perfect.

In amplifying this input signal, we'll keep it simple, making only a single pass through the
amplification.  We'll first consider the amplifier modelled by the sine curve of
Fig.~\ref{fig:linnonlinamp}.

If both notes are played with the same volume or amplitude, we create the power spectrum on the
left-hand side of Fig.~\ref{fig:2notesSine}.  This time we are focusing on the lower frequencies $<
1000$ Hz.  Here the harmonics have similar amplitudes creating a rich listening experience.

Upon amplifying this input signal with the nonlinear sine curve of Fig.~\ref{fig:linnonlinamp} we
obtain the power spectrum in the right-hand plot of Fig.~\ref{fig:2notesSine}. We observe the creation
of new notes by the amplifier: new notes that were not played by the musician.

The new note at 275 Hz has a large amplitude and will be prominent.  Relative to the A tonic note
the frequency ratio $275/110 = 5/2$.  Dividing by 2 to bring the tone down an octave, the ratio is
5/4.  Referring to Table \ref{tab:EqTempVsHarmonic} this is a major 3rd interval.  Thus our power
chord of the tonic and the perfect 5th has developed the properties of an A major chord through the
use of a distorted amplifier.  If you found yourself playing {\em Deep Purple}'s ``Smoke on the
Water'' on an acoustic guitar with full bar chords to make it sound good, now you know
why!\footnote{It's actually played by plucking two guitar notes simultaneously and letting the
amplifier do the rest.}

Similarly the new note at 385 Hz has a ratio to the tonic of $385/110 = 7/2$. Dividing by 2 to
bring the tone down an octave, the ratio is 7/4.  Referring to Table \ref{tab:EqTempVsHarmonic}
this is a harmonic minor 7th interval, the famous barbershop interval.  No wonder guitar amplifier
distortion sounds so glorious.  This process of filling in the gaps of the input power spectrum
continues to all higher harmonics.

\begin{figure}[tb]
  \includegraphics[width=0.49\textwidth]{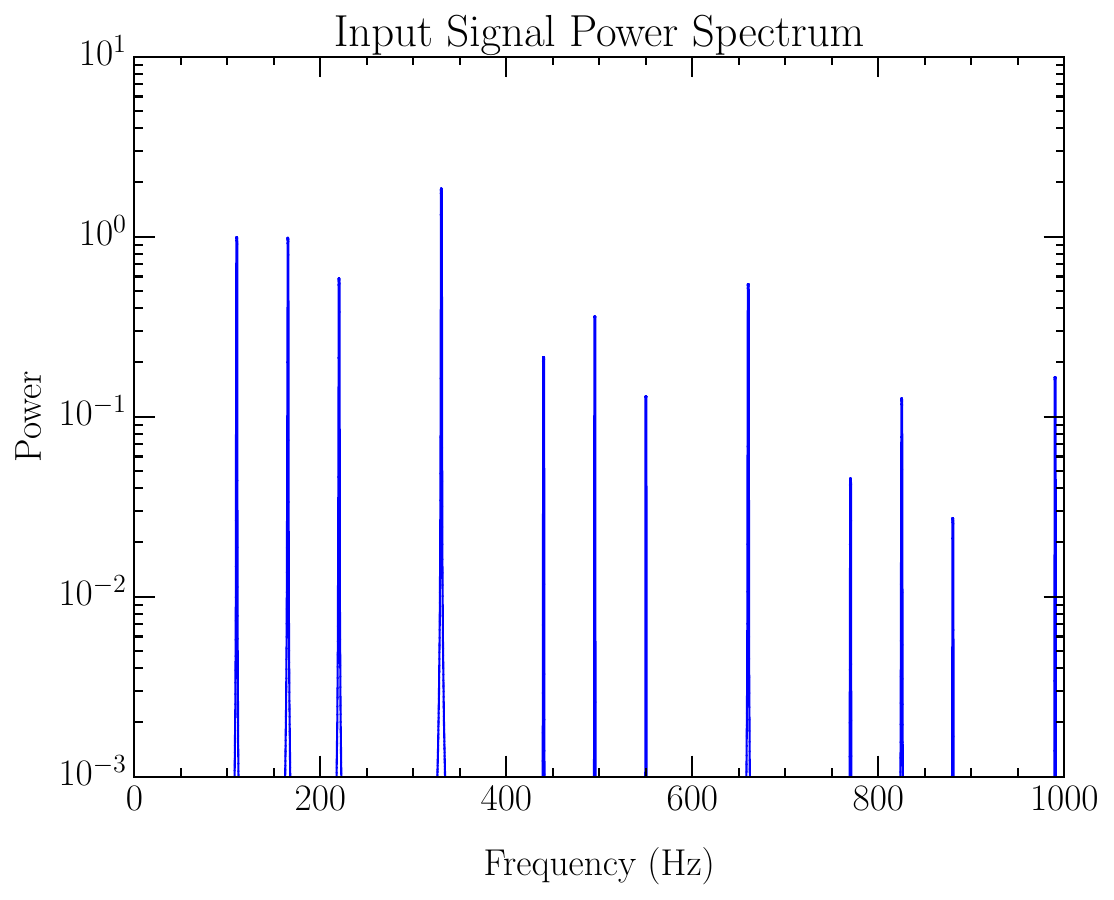}
  \includegraphics[width=0.49\textwidth]{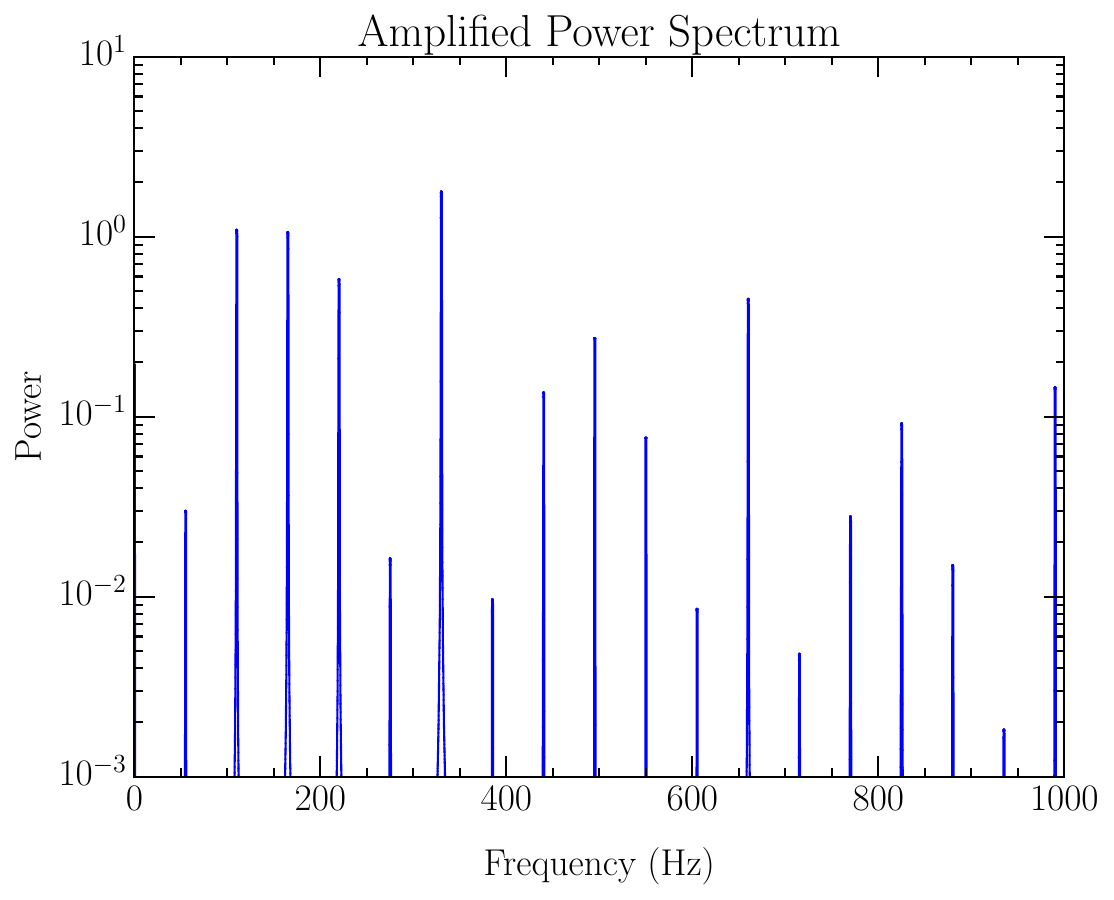}
  \caption{Power spectra for input (left) and output (right) wave forms after three passes of the
    nonlinear polynomial amplification curve.  This time a new tone at 55 Hz is created by the
    amplification.  This is in addition to the new tones created by sine amplification.  }
  \label{fig:2notesPoly3}
\end{figure}

Figure \ref{fig:2notesPoly3} illustrates the results for two passes on the nonlinear polynomial
amplification curve. Two passes are considered to generate a similar amount of distortion to that
produced in a single pass of the sine amplification. This time a new tone at 55 Hz is created by
the amplification.  This is in addition to the new tones created by sine amplification.  The ration
to the tonic at 110 Hz is 1/2.  Thus the inclusion of a term symmetric in amplification has
generated a new tone one octave down from the original fundamental.  And with the harmonic series
of integer multiples of 55 Hz complete in the output power spectrum, we will hear a new note an
octave lower than that played by the musician.  The A note at 55 Hz is even lower than the 6th E
string on the guitar.  You know it's going to sound good.

\section{Where do the new frequencies come from?}
\label{sec:trig}

\subsection{Nonlinear amplification}

When musicians drive their amplifiers hard, they can feed in two or three notes as input and find
that the amplifier adds tens (or hundreds) of new audible tones to the output. The power spectrum
now looks nothing like the original input! The new frequencies seem to have sprung up mysteriously,
but actually we find that they’re closely related to the input frequencies by some simple rules.

A handy way of breaking down the differences comes from separating the harmonics by odd and even
numbers. To pin down the source of odd and even sets of harmonics, it makes sense for us to look
towards the odd and even parts of the amplifier function.

A function like the nonlinear sine amplifier function in Figure \ref{fig:linnonlinamp} can be
written as a power series expansion. Expanding our function helps us to find out how the function's
nonlinear components are affecting the output frequencies. Labelling a few constants $k_1$, $k_2$
and $k_3$, the output voltage looks like this:
\begin{equation} 
V_{\rm out} = k_1\, V_{\rm in} + k_2\, V_{\rm in}^2 + k_3\, V_{\rm in}^3 + \cdots \, .
\label{eq:taylorexpandoutput}
\end{equation}
When the constant $k_1$ is the only non-zero constant in the expansion, then the amplifier is
linear, and its output depends linearly on the input according to the straight green line in Figure
\ref{fig:linnonlinamp}.

However, when the constants $k_2$, $k_3$ or higher are non-zero, no matter what value $k_1$ has,
the amplification function becomes curved. Consider the curve in Fig.~\ref{fig:linnonlinamp},
modelling a valve amplifier's response as a sine function. This function has a power series
expansion with $k_1 = 1$, $k_2 = 0$, and $k_3 = \frac{1}{3!}$, {\it i.e.}
\begin{equation} 
\sin(V_{\rm in}) = V_{\rm in} - \frac{V_{\rm in}^3}{3!} + \frac{V_{\rm in}^5}{5!} - \cdots \, .
\label{eq:sineseries}
\end{equation}
We notice here that a sine function only contains odd powers of the input. 
The polynomial curve of Fig.~\ref{fig:linnonlinamp} has $k_1 = 1$, $k_2 = a = 0.2$ and $k_3 = -a$
providing a source of even powers of the input.

\subsection{Odd and even distortion}

The design of the amplifier determines the values of the constants in
Eq.~(\ref{eq:taylorexpandoutput}).  Amplifiers are generally odd, as depicted in Figure
\ref{fig:linnonlinamp}, with some even components that are introduced to enhance the harmonic
content. Recall the generation of a new bass tone in Fig.~\ref{fig:2notesPoly3}.

The even terms introduce a difference in the amplification of positive versus negative input
voltages; they break the otherwise perfect anti-symmetry. For example, a positive even term in the
amplification function can enhance the amplification of positive voltages while suppressing the
amplification of negative voltages. This asymmetry can be regarded as a feature in valve
amplifiers.

Let us explore what the difference between even and odd amplification looks like when we expand the
output in terms of the constants in Equation~\ref{eq:taylorexpandoutput}. An odd amplifier will
have no even powers of the input voltage, so $k_2$ is zero, whereas an even amplifier will have no
odd powers, so $k_1$ and $k_3$ must both be zero. With a single input frequency, it's fairly
straightforward to see the effect that these constants have on our output.

\begin{figure}[tb]
	\includegraphics[width=0.32\textwidth]{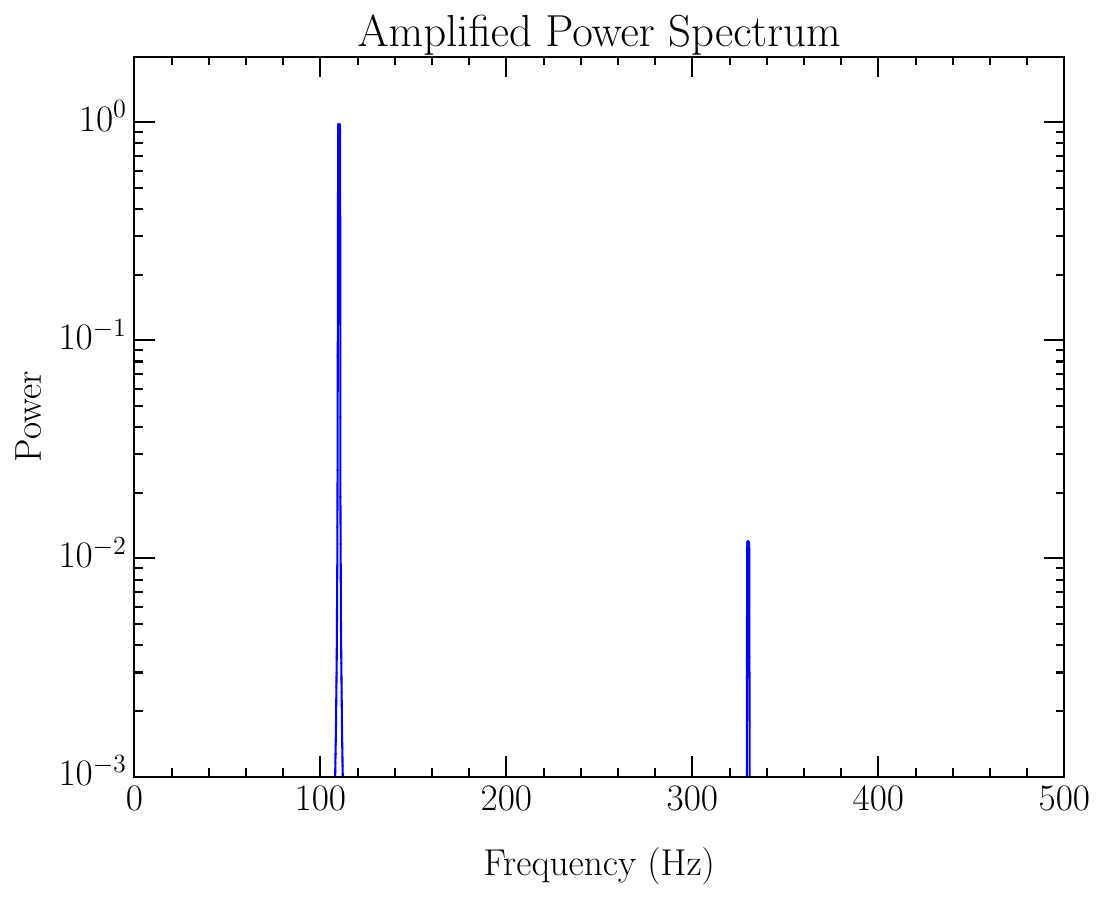}
	\includegraphics[width=0.32\textwidth]{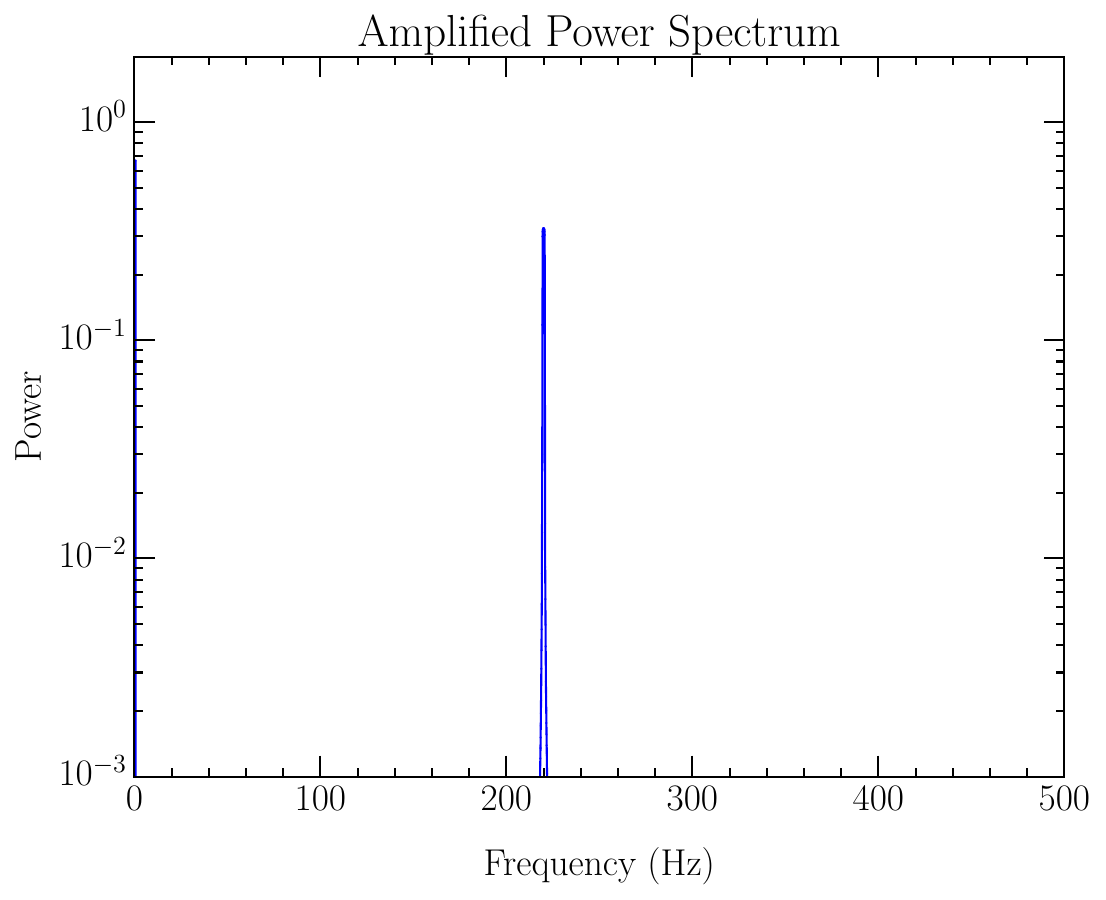}
	\includegraphics[width=0.32\textwidth]{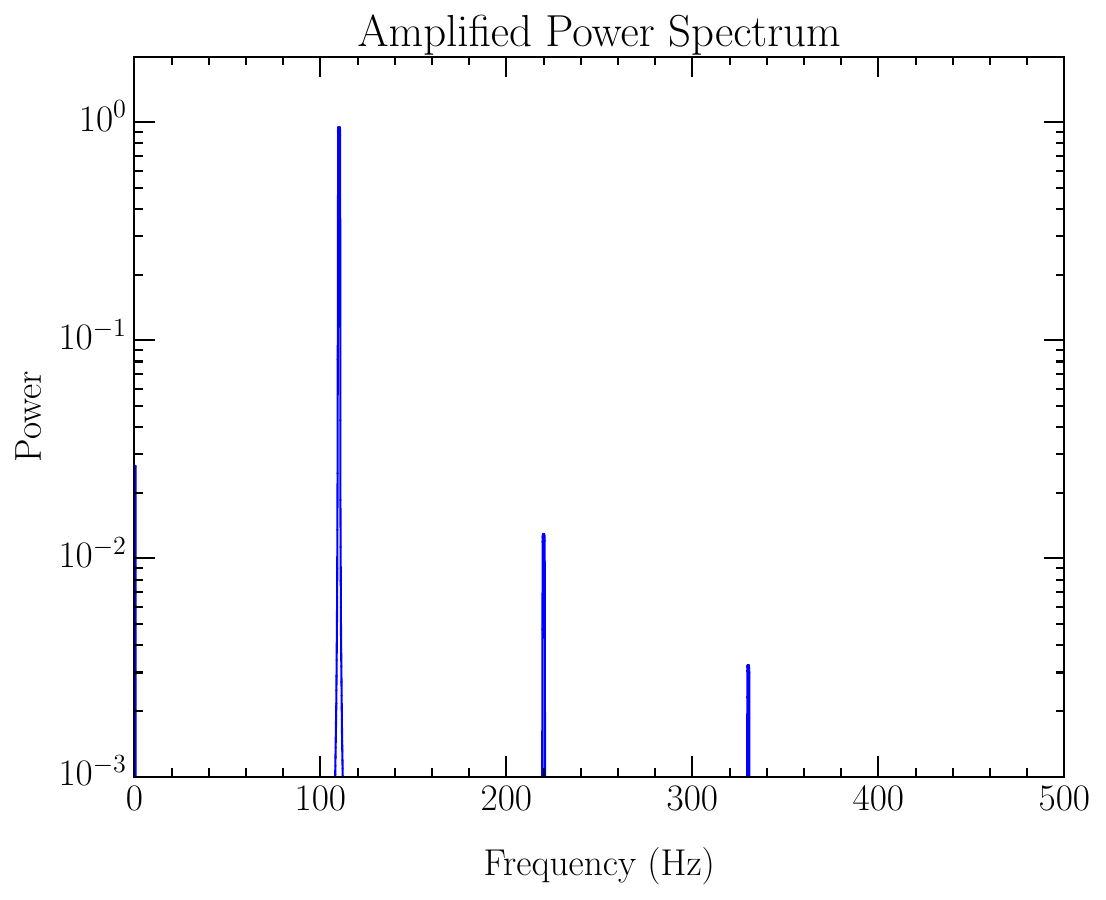}
	\caption{Power spectra for the output of an odd (left), even (middle), and polynomial
          (right) amplifiers driven by a single input tone of the A2 fundamental frequency at 110
          Hz corresponding to the open 5th string on a guitar.  Amplifiers are described in the text.}
	\label{fig:1tone}
\end{figure}

Figure \ref{fig:1tone} displays the power spectra created by amplifying a single A2 fundamental
tone at 110 Hz corresponding to the open 5th string on a guitar.  Results are displayed for odd,
even and mixed odd plus even amplifiers.

For the odd amplifier, we set $k_1 = 1$ and $k_3 = -1/3$ such that the slope of the amplification
curve is zero at $V_{\rm in} = 1$, similar to the sine function at $\sin(\pi/2)$.  
For the even amplifier we set $k_2 = +1/3$ in the spirit of matching the magnitudes of the
nonlinear coefficients as we did for the polynomial amplification curve.  The mixed amplifier is
the polynomial amplifier of Fig.~\ref{fig:linnonlinamp}.

We see that we still hear the input frequency very strongly when we put it through an odd
amplifier, but not when we use an even one.  Our odd amplifier generates only two tones and it's
tempting to think that it is the presence of two terms in the amplification function that results
in exactly two tones. The first tone remains at 110 Hz and the second tone is at 330 Hz, presumably
due to the presence of the cubic term in the amplification function.  Indeed, the odd sine
amplifier generates many harmonics, including the next harmonic at 550 Hz, five times the input
frequency, presumably due to the presence of a $V_{\rm in}^5$ in the Taylor-series expansion of the
sine function. This harmonic at 550 Hz is not generated by the odd amplifier limited to a cubic
term.  The even amplifier with a squared term only generates at tone at 220 Hz, double the input
frequency.

Pythagoras is back in our story again to give us clarity to how nonlinear amplification is
generating new tones. Despite how much our high-school friends complained about Pythagoras'
trigonometry being useless in real life, it's vital for understanding amplifier distortion in
modern music. We can predict exactly what frequencies will come out of an amplifier just by using
some simple trigonometric relations.

Let us  describe one input tone as a cosine function with an angular frequency of 1 for simplicity,
$V_{\rm in} = \cos(t)$, then our simple even amplifier produces an output with this frequency
\begin{equation}
V_{\rm out,\, even} = k_2\, \cos^2(t) \, .
\end{equation}
For our odd amplifier, we instead hear these frequencies: 
\begin{equation}
	V_{\rm out,\, odd} = k_1\, \cos(t) + k_3\, \cos^3(t)  \, .
\end{equation}
Now let's draw on some of Pythagoras' trigonometry relations
\begin{equation}
\cos^2(t) = \frac{1}{2} + \frac{1}{2} \, \cos(2t) \, ,
\label{eq:trigCos2}
\end{equation}
and
\begin{equation}
\cos^3(t) = \frac{3}{4}\, \cos(t) + \frac{1}{4} \, \cos(3t) \, .
\label{eq:trigCos3}
\end{equation}
In a power spectrum, $cos(2t)$ is double the input frequency, an octave higher than the input tone:
the second harmonic. The third harmonic comes from $cos(3t)$.  Thus the second harmonic only gets
generated by the {\em even} term of an amplifier and the third harmonic is dependent on the {\em
  odd} amplifier terms.

\subsection{Which sounds better?}

The debate over whether a signal with predominantly even or odd harmonics sounds more pleasant can
be hilariously divisive, to the extent that some people may worry about it and record their voices
to find out what type of harmonics their vocal chords naturally produce.  Many people report
preferring even harmonics to odd harmonics.

Referring back to Fig.~\ref{fig:firstpowerspectrum}, we see
the 2nd, 4th, and 8th harmonics are just octaves of the fundamental tone.  Only the 6th harmonic
brings in the new perfect 5th.  But this overtone is high in the spectrum with a smaller
amplitude. 

Odd harmonics are more interesting with the 3rd, 5th and 7th harmonics corresponding to the perfect
5th, major 3rd and harmonic minor 7th, a beautiful chord of four perfect intervals, but maybe that's
not what you're looking for in amplifying a single tone.

One might say that the even harmonics are ``more harmonic,'' because they contain more of the
lowest harmonics, since the second is lower than the third. However, the second harmonic is just
the octave interval, which is less interesting than the third harmonic, the origin of that powerful
perfect fifth.

Of course, in reality most amplifiers are not entirely odd or even, and their values of the
constants $k_1$, $k_2$, $k_3$ and so on are never exactly zero, although some may tend to amplify
either odd or even harmonics more than others. Valve or tube amplifiers tend to saturate
asymmetrically, and this adds more even harmonic frequencies, claimed to give their sound a more
pleasant harmonicity. If you believe this, it could be something to watch out for when buying an
amplifier.

\subsection{Two tones under nonlinear amplification}

The output frequencies we have just seen are all part of the harmonic spectrum of our starting
frequency, so of course we enjoy hearing them. Who doesn't like small integer ratios? However, all
the harmonics are integer multiples of the fundamental frequency, so we're still experiencing a
single note, but perhaps a littler richer than that due to changes in the shape of the harmonic
series as illustrated in Figs.~\ref{fig:1note} and \ref{fig:1notepoly}. 

But when the input contains multiple notes -- even just two -- the output is much richer as
illustrated in Fogs.~\ref{fig:2notesSine} and \ref{fig:2notesPoly3}.  This time new notes are
created by the amplifier; these new notes are not played by the musician.  But where do they come
from?  What is the physics that underlies this phenomena?

We can use the exact same approach as in the previous section to predict which notes we will hear
after amplification.  Because there is more than one tone, we'll see how the two tones come
together to create new tones or frequencies upon distorted amplification.  The new tones are
derived from the nonlinear amplification terms.  These take the sum of several input frequencies
and multiply that sum by itself several times, generating a lot of complicated cross-terms.

Let's try it. Let's consider two input tones with frequencies $f_1$ and $f_2$ and amplitudes $A_1$
and $A_2$, then the input voltage looks like 
\begin{equation}
V_{\rm in}(t) = A_1 \, \cos(f_1\, t) + A_2 \, \cos(f_2\, t) \, .
\end{equation}
After these are fed into the amplifier function, what new frequencies do we hear? Instead of
multiplying the same frequency by itself, the amplification will multiply the different frequencies
together. The tones that we hear are hidden inside the maths.

Let's consider the leading even nonlinear amplification term of our polynomial amplifier
\begin{eqnarray}
V_{\rm even}(t) &=& a\, \left ( A_1 \, \cos(f_1\, t) + A_2 \, \cos(f_2\, t) \right )^2 \, , 
\nonumber \\
&=& a\, \left ( A_1^2 \, \cos^2(f_1\, t) + 2\, A_1\,A_2 \, \cos(f_1\, t) \, \cos(f_2\, t) + A_2^2 \,
\cos^2(f_2\, t) \right ) \, . 
\label{eq:evenAmpCos}
\end{eqnarray}
Now let's use trigonometry relation
\begin{equation}
\cos(a)\, \cos(b) = \frac{1}{2} \left ( \cos(a-b) + \cos(a+b) \right ) \, .
\end{equation}
which reduces to our previous relation of Eq.~(\ref{eq:trigCos2}) for $a=b$.  Applying this to
Eq.~(\ref{eq:evenAmpCos}), and noting $\cos(a) = \cos(-a) = \cos(\vert a \vert)$, we discover
\begin{eqnarray}
V_{\rm even}(t) &=& \frac{a}{2}\, \left ( 
  A_1^2 \, \cos(2 f_1\, t) 
+ A_2^2 \, \cos(2 f_2\, t) 
+ A_1^2 + A_2^2 \right . \nonumber \\
&& \quad \left .
+ 2\, A_1\,A_2 \, \cos(\vert f_1-f_2 \vert\, t)
+ 2\, A_1\,A_2 \, \cos((f_1+f_2)\, t)
\right ) \, . 
\label{eq:evenAmpNewFreq}
\end{eqnarray}
The first two terms are familiar with integer multiples of the original frequencies.  But the last
two terms are new, and louder.  The amplitude is twice that of the familiar terms for $A_1 = A_2$,
like in ``Smoke on the Water'' where two strings are plucked evenly and simultaneously.  Two new
frequencies appear at sums and differences of the original frequencies.  In summary, the even
quadratic amplification function generates four frequencies
\begin{equation}
\vert 2 f_1   \vert \, , \quad
\vert f_1 - f_2 \vert \, , \quad
\vert f_1 + f_2 \vert \, . \quad
\vert 2 f_2   \vert \, ,
\end{equation}
with the cross terms twice the amplitude of the squared terms.  

\begin{figure}[tb]
	\includegraphics[width=0.48\textwidth]{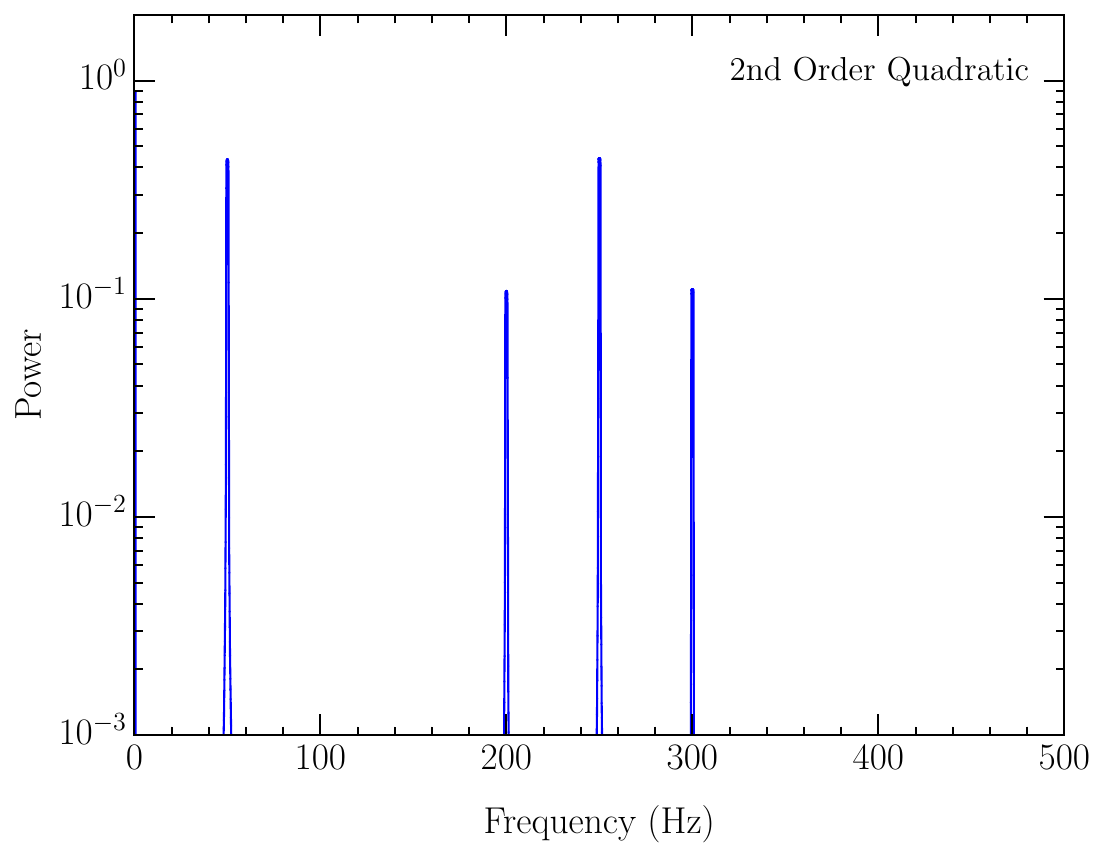}\quad
	\includegraphics[width=0.48\textwidth]{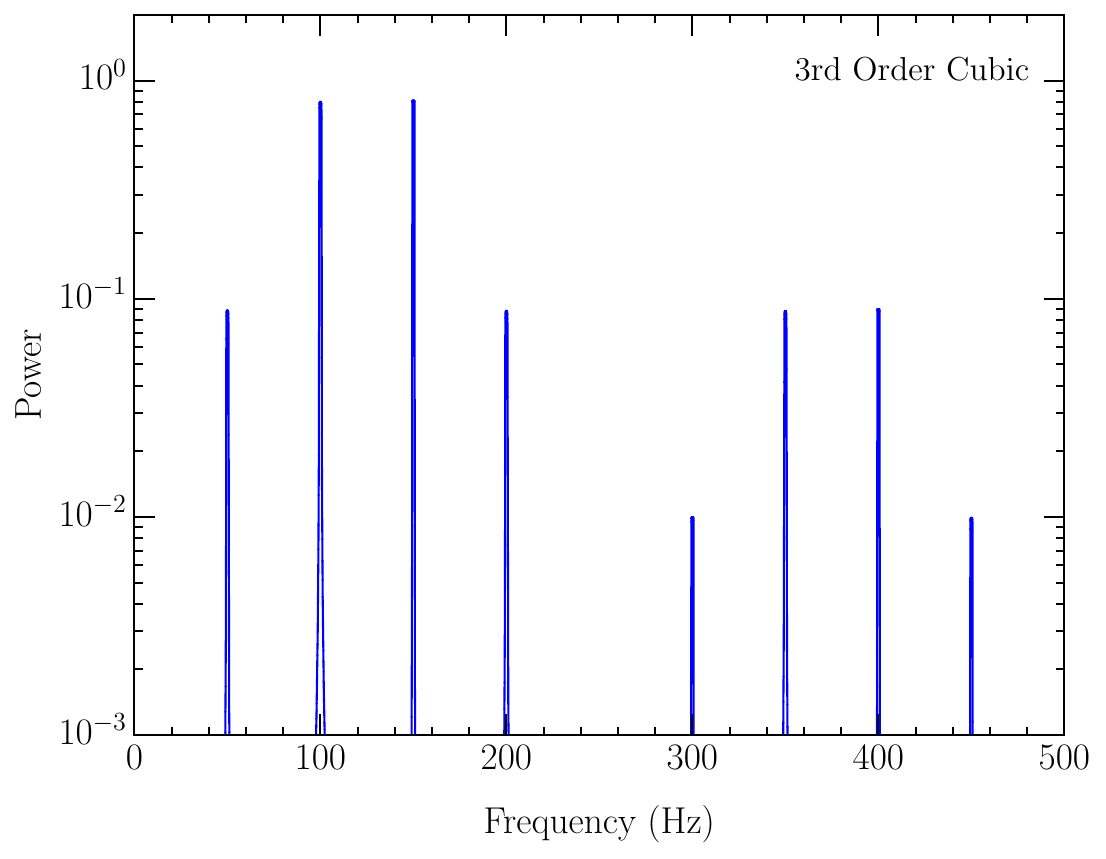}\\[6pt]
	\includegraphics[width=0.48\textwidth]{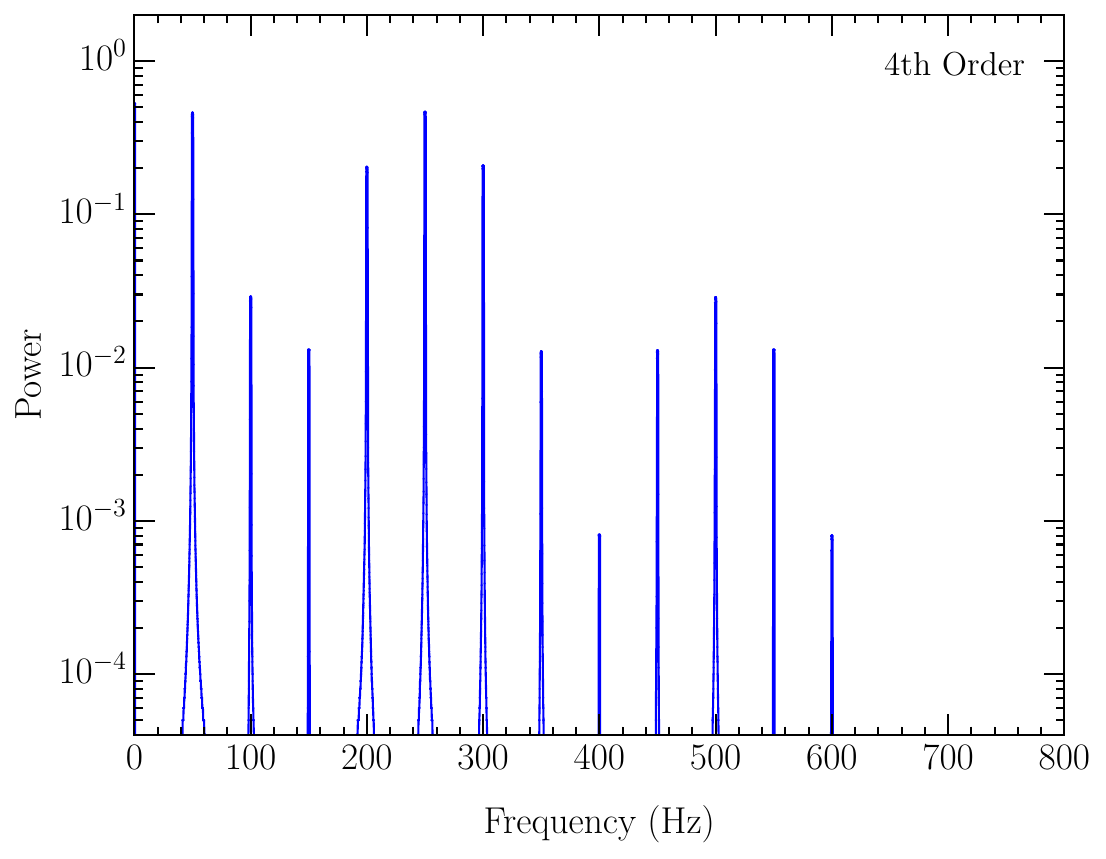}\quad
	\includegraphics[width=0.48\textwidth]{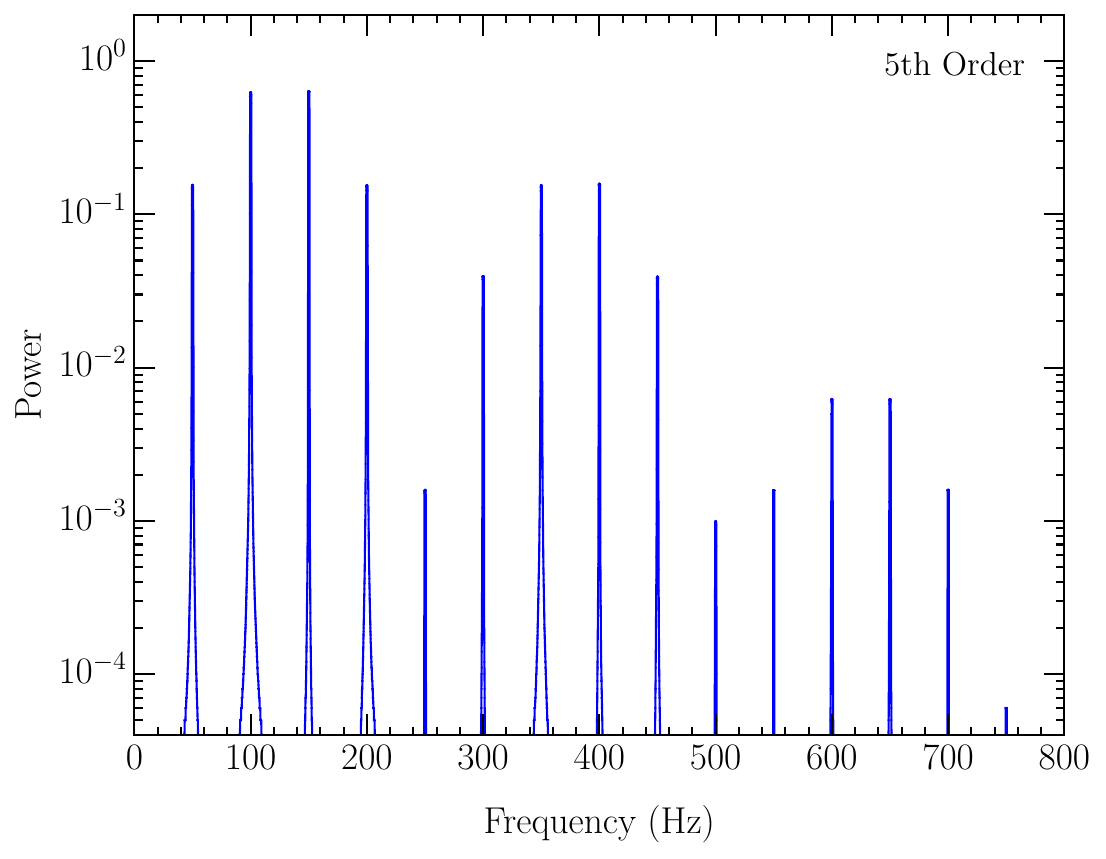}
	\caption{Power spectra generated by various nonlinear terms in a Taylor expansion of a
          distorting amplifier function. The input is composed of only two tones, with $f_1 =
          100\,$Hz and $f_2 = 100\times 3/2 = 150\,$Hz, an interval of a perfect 5th. Taylor
          expansion terms include $V_{\rm in}^2$ (upper left), $V_{\rm in}^3$ (upper right),
          $V_{\rm in}^4$ (lower left), and $V_{\rm in}^5$ (lower right).  The new frequencies are
          generated from sums and differences of the two input frequencies, a process called
          inter-modulation distortion. The $V_{\rm in}^2$ and $V_{\rm in}^3$ terms are described in
          detail in the text.  In comparison to the harmonic series of a single note, the output is
          much more complicated providing a rich listening experience. }
	\label{fig:TaylorTerms}
\end{figure}

For example, suppose $f_1 = 100\,$Hz and $f_2 = 100\times 3/2 = 150\,$Hz with equal amplitudes.
Then the new frequencies are $50,\ 200,\ 250,\ \mbox{and}\ 300\,$Hz, with the 50 and 250 Hz
amplitudes twice that of the 200 and 300 Hz amplitudes.  The new low tone is an octave down from
the fundamental of 100 Hz and the frequency ratio $250/100 = 5/2$ is an octave above the first
major third creating a major chord.  The power spectrum for this scenario is illustrated in the
upper-left plot of Fig.~\ref{fig:TaylorTerms}.

Now we understand the origin of that new tone in Fig.~\ref{fig:2notesPoly3} an octave below the
original fundamental tone played by the musician.  Recall the two notes played have the fundamental
frequencies of $f_1 = 110\,$Hz, the open 5th string, and the harmonically tuned E played at the 2nd
fret on the 4th string with $f_2 = 110 \times 3 / 2 = 165\,$Hz.  The low tone is $\vert f_1 - f_2
\vert = 55\,$Hz, half of $f_1$ and therefore an octave lower.  Thus the selection of the ``power
chord,'' the tonic and the perfect fifth, is vital to getting this octave lower tone.  It will not
appear for other intervals.  It would seem the name ``power chord'' is justified by the physics.

Now let's consider the cubic term of the odd amplifier
\begin{eqnarray}
V_{\rm cubic}(t) &=& -a\, \left ( A_1 \, \cos(f_1\, t) + A_2 \, \cos(f_2\, t) \right )^3 \, , 
\nonumber \\
&=& -a\, \left ( 
  A_1^3 \, \cos^3(f_1\, t) 
+ 3\, A_1^2\,A_2 \, \cos^2(f_1\, t) \, \cos(f_2\, t) 
\right . \nonumber \\
&& \qquad \left . 
+ 3\, A_1\,A_2^2 \, \cos(f_1\, t) \, \cos^2(f_2\, t) 
+ A_2^3 \, \cos^3(f_2\, t) 
\right ) \, . 
\label{eq:cubicAmpCos}
\end{eqnarray}
We've seen the cubic terms before creating a new tone at $3  f_1$ and $3 f_2$.  As integer
multiples of the fundamental frequency, there are not as interesting as the cross terms in
Eq.~(\ref{eq:cubicAmpCos}).  These terms are even louder than the even quadratic term considered
earlier with a coefficient of 3.  This time we'll need the general relation
\begin{equation}
\cos(a) \, \cos(b) \, \cos(c) = \frac{1}{4} \cos(a + b + c) + \frac{1}{4} \cos(a + b - c) + \frac{1}{4} \cos(a - b + c) + \frac{1}{4} \cos(a - b - c).
\end{equation}
Thus the new frequencies generated are
\begin{equation}
\vert 3 f_1       \vert \, , \quad
\vert   f_1       \vert \, , \quad
\vert 2 f_1 - f_2 \vert \, , \quad
\vert 2 f_2 - f_1 \vert \, , \quad
\vert 2 f_1 + f_2 \vert \, , \quad
\vert 2 f_2 + f_1 \vert \, . \quad
\vert   f_2       \vert \, , \quad
\vert 3 f_2       \vert \, ,
\end{equation}
a total of eight frequencies generated from the original two.  However the amplitudes of these
frequencies vary. The four combinations of frequencies $f_1$ and $f_2$ have three times the
amplitude of the $3 f$ frequencies. All four terms of Eq.~(\ref{eq:cubicAmpCos}) generate the
original frequencies $f_1$ and $f_2$ and these frequencies have nine times the amplitude of the $3
f$ frequencies.  In the power spectrum these amplitudes are squared such that the original
frequencies are 81 times louder than the $3 f$ frequencies, and 9 times louder then the frequency
combinations.  
This contrasts the even amplifier where the original frequencies are not produced.  In this case
the new frequencies can be enhanced to create a richer listening experience.

To understand this more clearly, again suppose $f_1 = 100\,$Hz and $f_2 = 100\times 3/2 = 150\,$Hz.
Then the new frequency combinations are $50,\ 200,\ 350,\ \mbox{and}\ 400\,$Hz and are complemented
by the lower amplitude $3 f = 300$ and $450\,$Hz frequencies.  Again the new low tone is an octave
down from the fundamental of 100 Hz and the frequency ratio $350/100 = 7/2$ is an octave above the
first barbershop harmonic minor seventh. The power spectrum for this scenario is illustrated in the
upper-right plot of Fig.~\ref{fig:TaylorTerms}.

Thus, we have learned that odd amplifiers do indeed create a low tone an octave down from the
fundamental frequency of the tonic for a power chord. However the reinforcement of the original
frequencies renders its contribution to be small relative to the original notes.  Indeed it is
there in the power spectrum, but it's too faint to appear in Fig.~\ref{fig:2notesSine} as the power
in the 55 Hz frequency is below $10^{-3}$.

Thus, the amplifier's output contains an eclectic mix of funky new combinations of the two input
frequencies. And we now understand that their presence actually derives easily from well-known
trigonometric identities. They provide a wonderfully straightforward way of seeing how input
frequencies combine when fed through a distorted amplifier. It happens whenever we feed in two or
more different frequencies; the amplifier produces all their sums and differences!  It can become
very complicated.

For example, the lower two plots of Fig.~\ref{fig:TaylorTerms} illustrate the increasingly complex
spectra generated from the two tones at 100 and 150 Hz in higher fourth-order $V_{\rm in}^4$
(lower-left) and fifth-order $V_{\rm in}^5$ (lower-right) terms.  The latter is the next term in
the Taylor expansion if the sine amplifier function.  Note how the even second- and fourth-order
terms are good at generating a new tone an octave below the fundamental frequency of the tonic of
100 Hz, with an amplitude much larger then the original input frequencies.  These even terms act to
break the perfect anti-symmetry of an amplifier and often have a presence in valve amplifiers.

The important point is that the new frequencies are not just the integer harmonics of the
frequencies we started with. The sums and differences are a signature of an effect called {\em
  Inter-modulation Distortion} (IMD) that tampers with the sound we want to make louder, distorting
what we thought we would hear into something else. Even though the new frequencies are not
harmonic, they can add interesting flair, whether it be the main point of focus in a distorted
amplifier or the subtle addition of frequencies in a studio processor.

\section{Problems with equal temperament}

Now we can see why the deviation of equal temperament from harmonic tuning is so problematic in a
highly distorted amplifier; when the input frequencies are a little out of tune, the sums and
differences of their frequencies will be even further out of tune. We saw that the Major 3rd is off
by 13.7 cents, which really makes a mess of the amplifier's output signal. Although the difference
can be subtle before amplifying, we surely hear that deviation in the output.

To hear the difference, compare the equal temperament and harmonic tuning sound samples available
at Ref.~\cite{PhysicsOfMusic}. There you will find links to mp3 files containing a combination of
the tonic note with the major 3rd, perfect 4th, and perfect 5th intervals of the scale in both equal
temperament and harmonic just tuning. Both clean undistorted notes, and valve distorted notes are
available.

So just how bad can the power spectrum for equal temperament look?  Here we show results for the
problematic interval of the tonic and the major 3rd of a scale.  Such intervals are common in
modern music, so we must handle them with care.

\begin{figure}[tb]
  \includegraphics[width=0.48\textwidth]{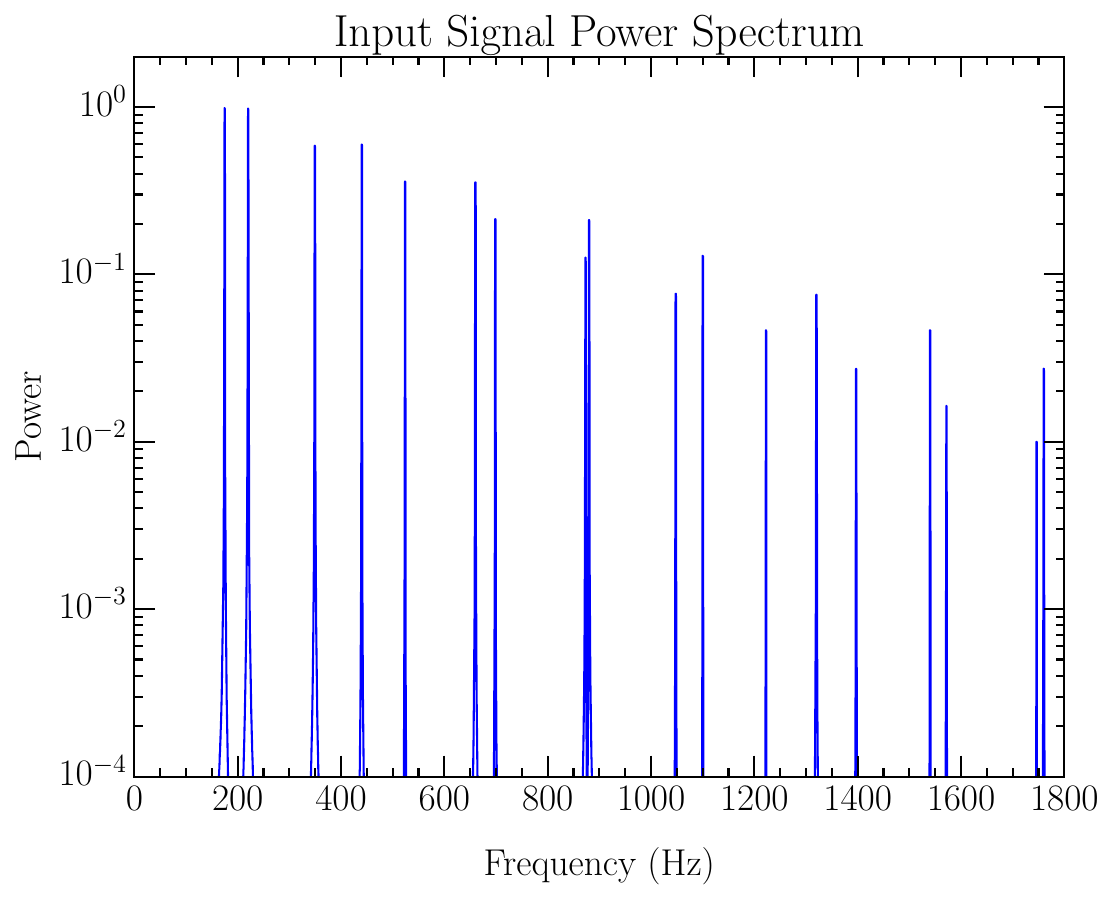} \quad
  \includegraphics[width=0.48\textwidth]{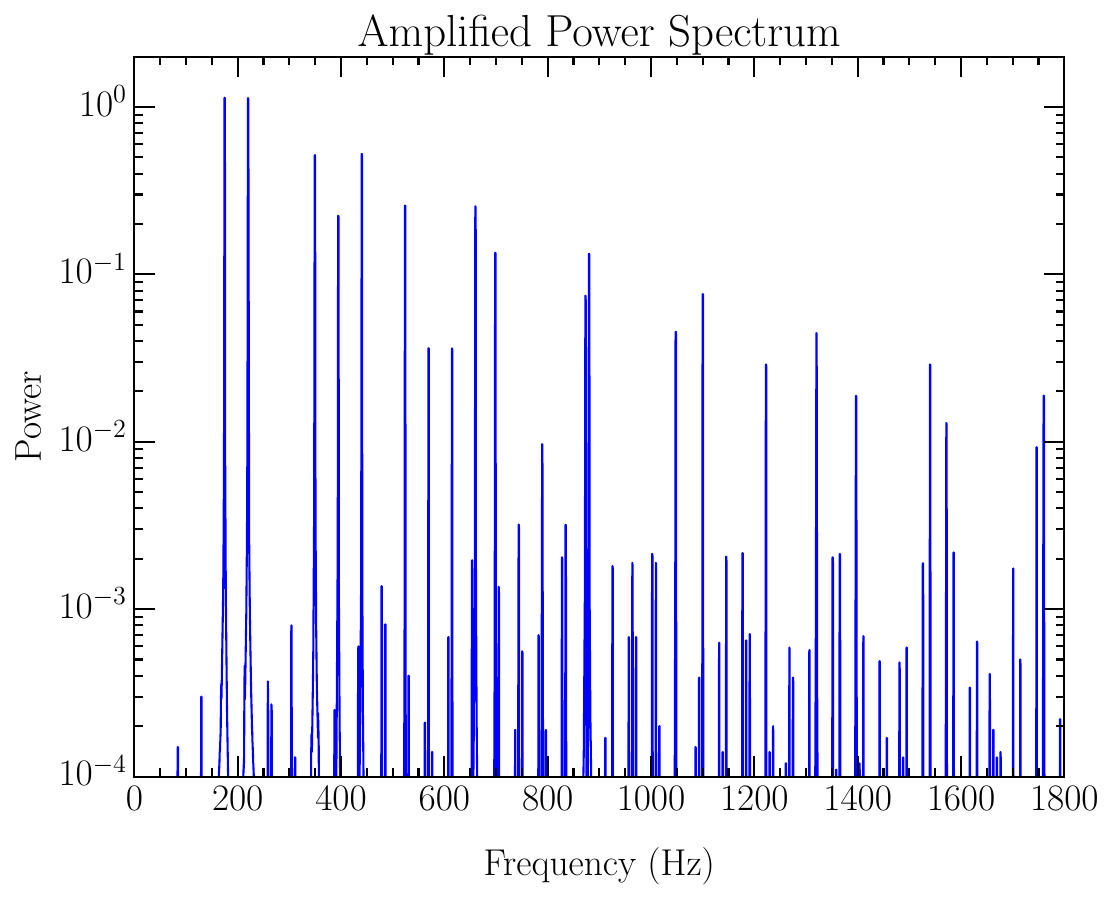}\\[6pt]
  \includegraphics[width=0.48\textwidth]{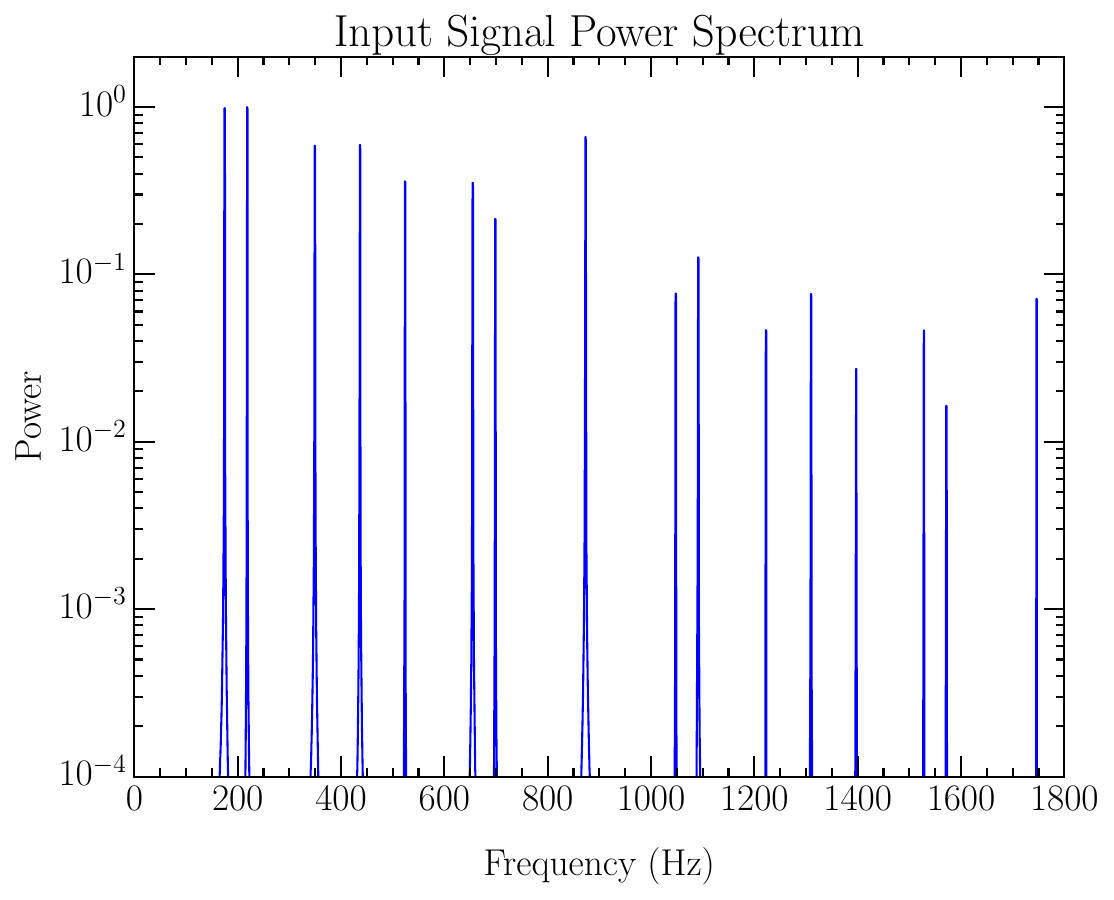} \quad
  \includegraphics[width=0.48\textwidth]{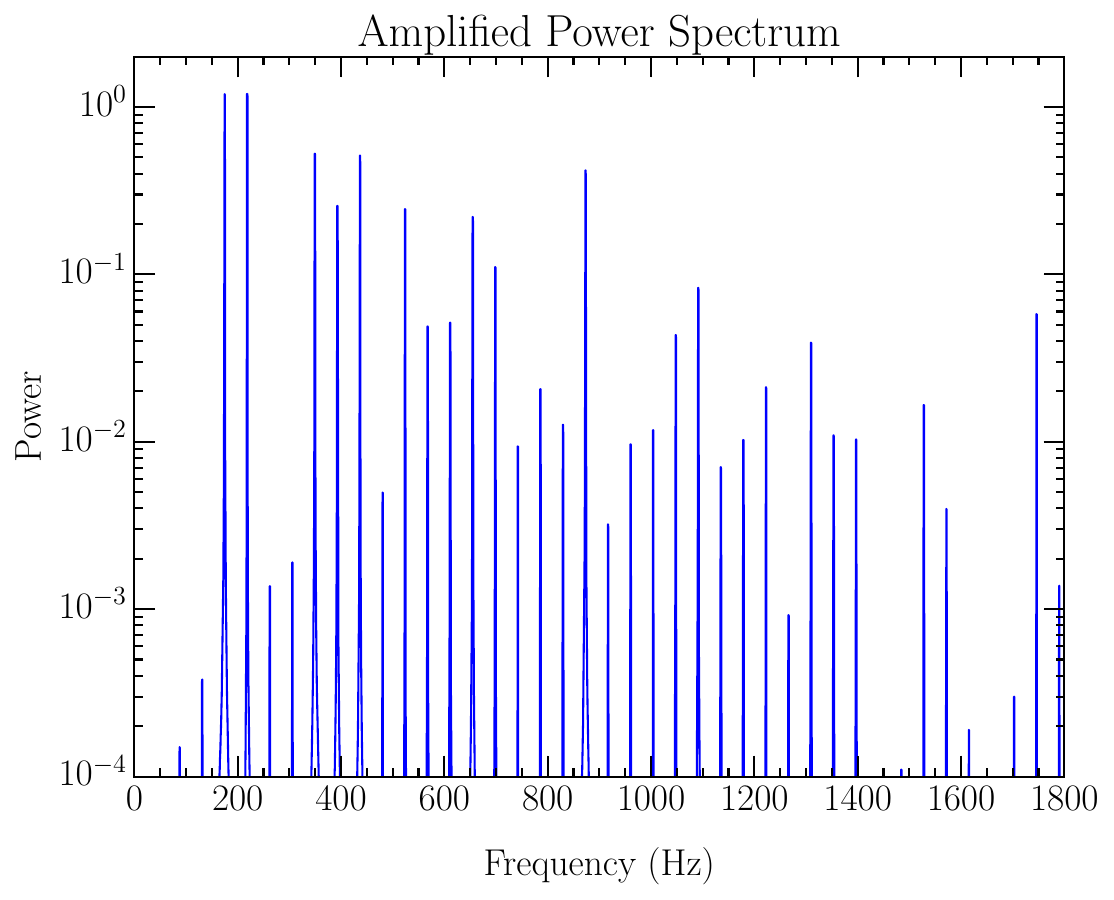}
  \caption{A comparison of the power spectra for two input signal notes at an interval of a major
    3rd in two different tuning schemes.  The upper plots illustrate the power spectrum for
    standard guitar-tuner-style equal temperament tuning where the major 3rd is sharp by 13.7
    cents, while the lower plots show the same information for by-ear harmonic just tuning. The
    irregular spacing of the equal temperament tuning creates a poor listening experience,
    especially after amplification by a three-gain-stage amplifier modelled by the sine function of
    Fig.~\ref{fig:linnonlinamp}. }
  \label{fig:3rdInterval}
\end{figure}

If we compare the top and bottom rows of Figure~\ref{fig:3rdInterval}, we see the difference
between equal temperament tuning (top row) and harmonic just tuning (bottom row). Just is in the
aforementioned mp3 files, we are considering an F on the fourth string of a guitar at fret 3 and
the major 3rd, A, on the third string at the 2nd fret with equal amplitudes.

In the upper left plot, the equal temperament tuning of the input signal contains harmonic
frequencies from each note that sit close to each other but are slightly different.  In some cases
these pairs may appear as thick lines, but these are actually two or more peaks that almost
coincide. Note the two harmonics slightly out of tune at approximately 875 Hz, 1510 Hz and 1,750
Hz. These frequencies have a difference that's so small that they {\em beat} strongly with an
unpleasant sound. The just tuning in the bottom left sounds pleasant by comparison.

The power spectra in the left-hand column of Fig.~\ref{fig:3rdInterval} are input to a nonlinear
amplifier applying three gain stages of the sine amplifier function. The output for each tuning is
illustrated in the right-hand column.

The problems with equal temperament are already apparent in the input signal, even without an
amplifier. After amplifying the equal-temperament input, we find in the top-right plot of
Fig.~\ref{fig:3rdInterval} a complicated power spectrum of bizarre frequencies that deviate from
our favourite pleasant regular intervals. The subtle differences in the frequencies means they
don't stack up as in the lower plot for just tuning. However, there is a great deal of power in the
spectrum scattered across many close frequencies creating an unpleasant sound with terrible
beating. This dissonant sound contrasts the lower-right plot for harmonic just tuning.  One observes
a rich spectrum of perfectly spaced harmonics indicating a most enjoyable listening experience.

The equal-temperament amplified power spectrum is populated with many side-by-side frequencies
because the problem already existed in the input signal.  Equal temperament creates errors in the
alignment of the harmonic frequencies that show as small spacings between some of the input
harmonic frequencies. From our
knowledge of nonlinear amplification, we can deduce that many of the amplified frequencies on the
right should appear at intervals of that small difference between the beating frequencies in the
left hand plot. Not only do we still have the beating frequencies in the output, but the beating
becomes much worse. The amplifier adds a cacophony of new frequencies that also beat with each
other because they are similar but dreadfully different.  The lesson:
\begin{center}
	\noindent
	\framebox[0.95\textwidth]{
		\vspace*{3mm}
		\begin{minipage}{0.9\textwidth}
			\vspace*{3pt}
			\centerline{\normalfont \bfseries Use your ears to get the tuning right for the song you are playing.}
			\vspace*{1pt}
		\end{minipage}
		\vspace*{3mm}}
\end{center}
\vspace{6pt}
You might need another guitar.  Go on, get that Certificate \cite{Certificate}.

\section{A phantom note}
\label{sec:phantom}

Such a complicated blend of frequencies begs the question: how do we perceive this
sound? We saw earlier that it's normal for us to perceive the entire harmonic spectrum of a note at
the equivalent of the fundamental, or lowest frequency: further proof that our brains are excellent
at processing a lot of information simultaneously. If we can process so many frequencies in a note,
then how do we process the frequencies produced by amplifier distortion?

It turns out that we can consider amplifier distortion to understand even better the remarkable
phenomenon of the ear-brain processing pathway. We again relate our understanding of the way we
appreciate amplified music back to -- you guessed it -- harmonics. Since amplifiers produce sums
and differences of the same frequency interval, we expect their output to contain a number of
overtones that match at least one harmonic series. However, when the amplifier does not produce all
the harmonics in that series, and could even be missing the important fundamental frequency, how do
we perceive the sound?

Consider first the scenario where all frequencies are present except the lowest, so our output
contains all harmonics starting from the octave above the fundamental. We might deduce that our
ear-brain system causes us to hear the lowest frequency that is present -- in this case, the second
harmonic. However, logical as this may seem, we still perceive the fundamental frequency -- even
though it is not actually being played.

A few reasons were proposed to explain this {\em missing fundamental} phenomenon. It was thought
that the ear-brain system acts as a \textit{second} nonlinear amplifier, further distorting the
sound we hear where one of the additional distortion frequencies is the missing fundamental. If
this is true, then the fundamental is physically present inside our ears: taking a look deep enough
into our inner ears, we should find a membrane vibrating at that frequency.

Our inner ear is indeed a nonlinear amplifier, adding new distortion frequencies just like an
electronic amplifier. To check your own ears for non-linearity, try the tests on this page
\cite{stephane_nonlinear_tests}. This is impressive in itself, considering the complicated
frequencies that must come out of \textit{two} nonlinear amplifiers. However, studies have revealed
systems where the inner ear's distortion frequencies do not include the missing fundamental, and
yet we still hear it. Speculations were abundant throughout this problem's history, when
interference beat notes, Tartini tones, the missing fundamental, combination tones, residue pitch,
virtual tones and difference tones were all terms on offer, resulting in several fascinating
discoveries in psychoacoustics \cite{10.1121_1.1918360, pressnitzer:hal-01105579,
  10.1121_1.1914648}.

Somewhere in the ear-brain system, whether in the inner-ear or through pitch-sensitive neurons in
the brain, the missing fundamental appears. The effect is powerful enough that it often escapes our
notice when applied in real life; for example, mobile phones are typically not capable of vibrating
at low enough frequencies to replicate low-pitched voices, so phones are designed to play the
harmonic series of those pitches without their fundamental. Instead of trying to vibrate, they wait
for the ear-brain system to fill in the rest.

Consider now the scenario where more than just the fundamental frequency is missing. What if only a
few harmonics are left in the series? What if we start removing the second, third, even the fourth
harmonic? This is a typical scenario for amplifiers, where the sum and difference frequencies of
distortion are only filling a limited number of series tones.

The bizarre answer is that indeed, we still hear the missing fundamental -- although within
limits. As long as the amplitudes of the remaining harmonics are large enough and the difference
between these harmonics is still equal to the fundamental frequency, then with enough harmonics
present we will interpret a much lower pitch than is playing. Eventually, the phenomenon breaks
down and we're likely to hear nothing but a jumble of higher tones, but the breakdown occurs after
removing more frequencies than one might expect.

The missing fundamental effect lends us the freedom to generate low sounds that might be
unreachable otherwise, which is pretty neat. If a nonlinear amplifier produces distortion aligned
with the harmonic series, then it should be useful for generating low sounds. All we should need to
do is feed an amplifier with an input that contains two frequencies, spaced apart by a difference
equal to the low note we're looking for. The distortion plus our ear-brain system should fill in
the rest. This is a great way to produce low frequencies that are difficult to play on an
instrument.

\begin{figure}[tb]
  \begin{center}
  \includegraphics[width=0.32\textwidth]{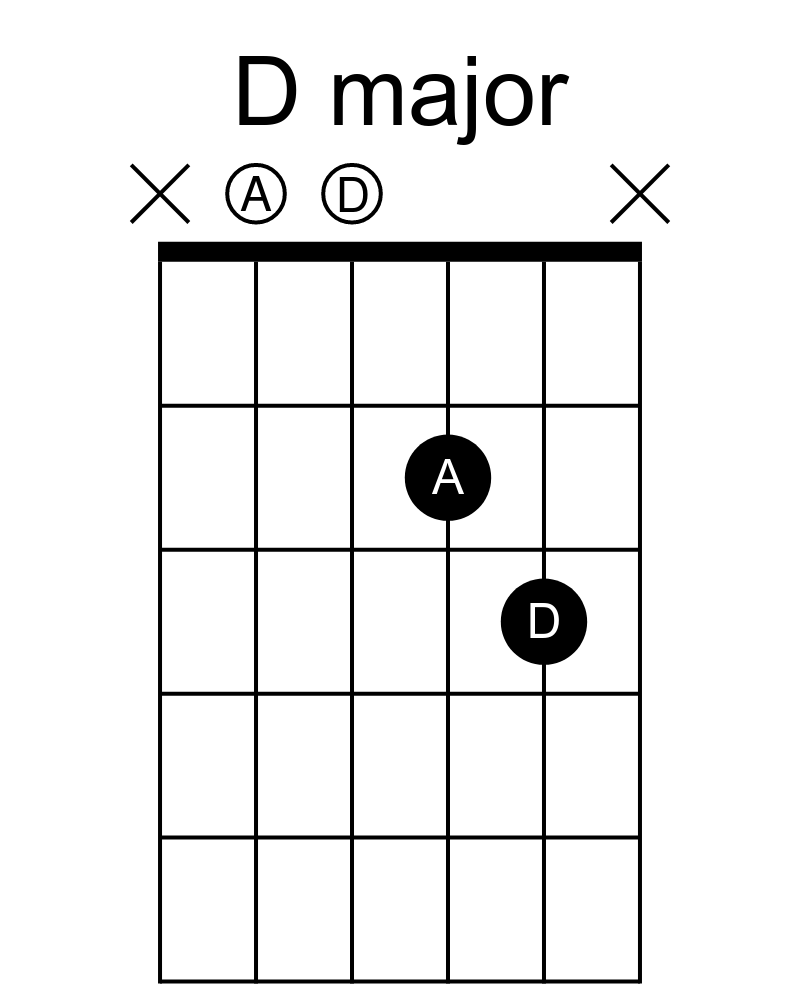}
  \quad
  \includegraphics[width=0.32\textwidth]{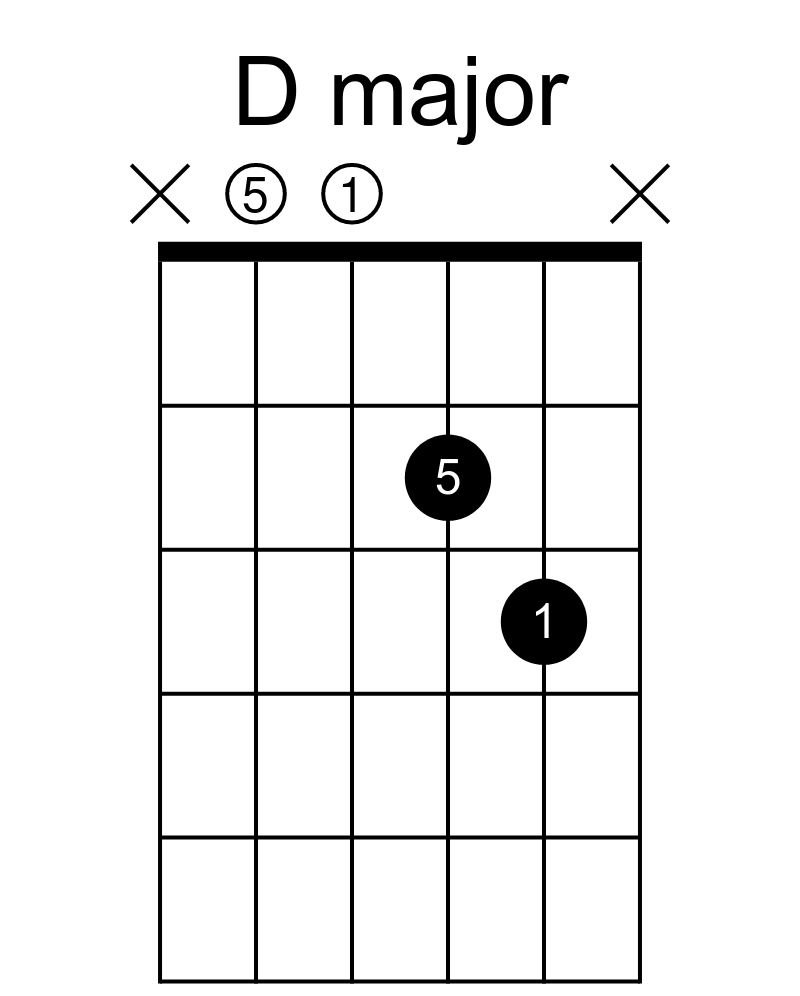}
  \end{center}	
  \caption{ Chord chart for a D-major power chord on a 6-string guitar in standard tuning.  Note
    names (left) and note intervals from the D tonic (right) are indicated. {\large $\times$}
    denotes the muting of the sixth and first strings. }
  \label{fig:Dchord}
\end{figure}

The classic case for guitar is the D major chord played on a 6-string guitar in standard tuning.
Here the 6th string is an E at 75 Hz but we're seeking a D at 73 Hz.  Sure you could do a drop-D
tuning, but that's cheating. The power chord is played on the four inside strings starting with A
$110\,$Hz on the 5th string.  In harmonic just tuning the D on the 4h string is
$146.\periodfl{6}\,$Hz, A $220\,$Hz on the third string, and D $293.\periodfl{3}\,$Hz on the 2nd
string.  The chord is depicted in Figure \ref{fig:Dchord}.  The idea is that with enough
distortion, we'll fill out the harmonic series for the D at $73.\periodfl{3}\,$Hz such that we can
hear it, even if it is not played.

To see how this works, we'll set up the problem for a D tuned a little sharp at 75 Hz to make it
easier to track the harmonics.  In Just tuning the A is at 3/2 times 75 Hz = 112.5, the next D is
at 150 Hz and the next A is at 225 Hz.  we're looking for a harmonic series with 75 Hz increments,
but not from a fundamental of 75 Hz, rather we're starting from the D played at 150 Hz.  To be
clear we are looking for harmonic strength at 150, 225, 300, 375, 450, 525, 600, 675,\ldots.

\begin{figure}[tb]
  \includegraphics[width=0.48\textwidth]{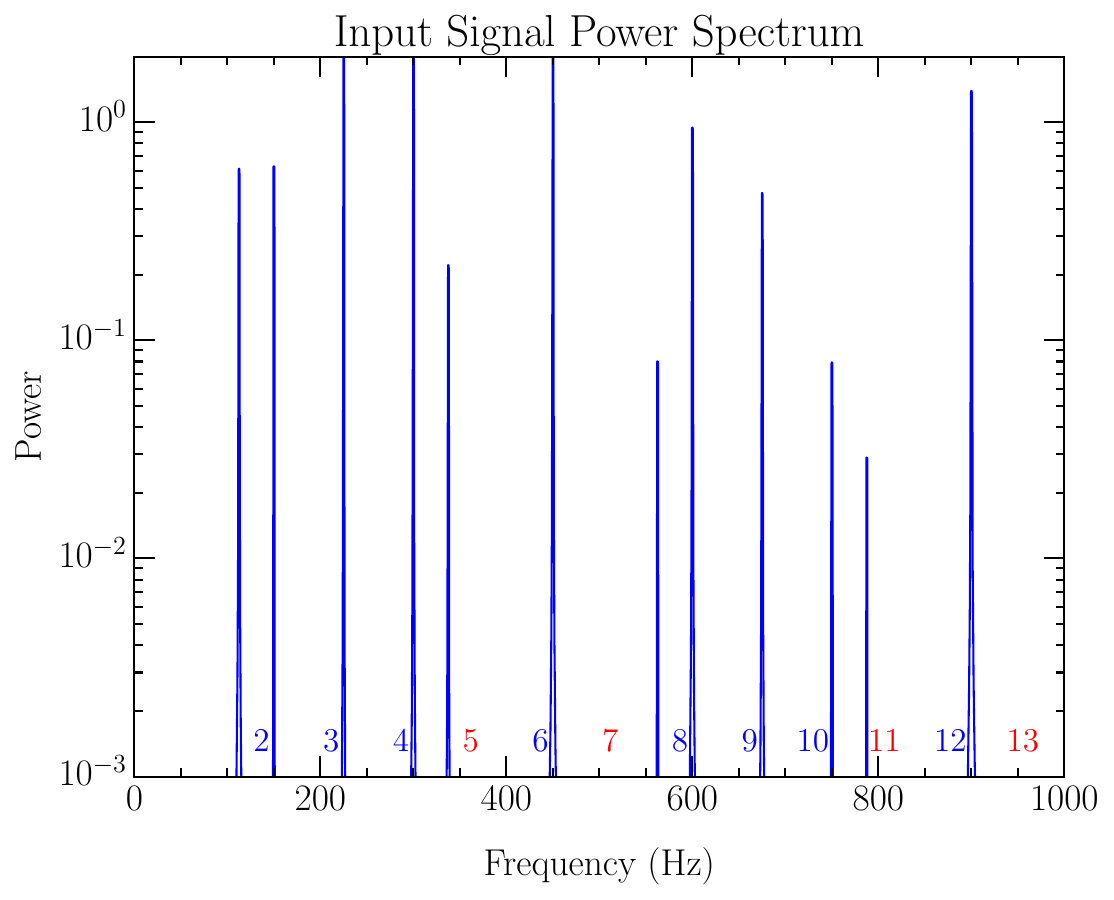} \quad
  \includegraphics[width=0.48\textwidth]{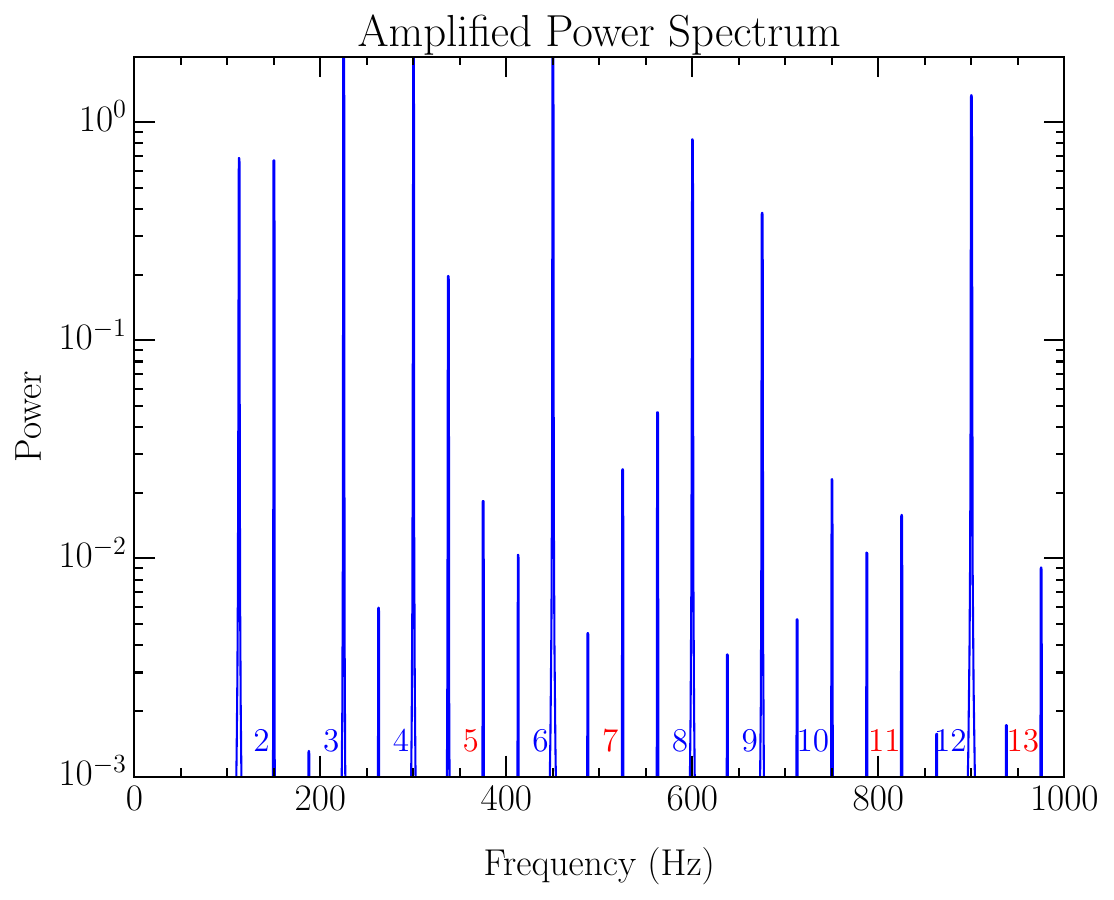}
  \caption{Input power spectrum (left) excited by playing a D-major chord as described in the text.
    To make it easy to track the positions of the harmonics, the D played on the open 4th string of
    a guitar is tuned a little sharp at 150 Hz.  Other notes are harmonically tuned relative to D
    150 Hz. The harmonic series of the missing fundamental at 75 Hz is numbered to the left of the
    spectral strength with missing strength indicated in red.  The power spectrum generated by the
    nonlinear sine amplification function of Fig.~\ref{fig:linnonlinamp} is illustrated in the
    right-hand plot.  Inter-modulation distortion has filled out the harmonic spectrum such that
    there is significant strength in the harmonic spectrum at integer multiples of 75 Hz commencing
    at 150 Hz.  Even though there is negligible strength at 75 Hz, we hear the missing
    fundamental.}
  \label{fig:4notes}
\end{figure}

The left-hand plot of Fig.~\ref{fig:4notes} illustrates the input frequencies excited by playing
the D-major chord as described above in the left-hand plot.  The frequencies include the harmonics
naturally excited by picking a string.  We see we're well on the way to a full harmonic series of
integer multiples of the missing fundamental at 75 Hz.  These frequencies are numbered 2 through 13
in the plot.  Strength at 150, 225, 300, 450, 600 and 675 Hz is already present but strength at
375, 525, 825, and 975 Hz is absent.  Our ears are clever and we don't hear a low D at 75 Hz.

But after one pass of the nonlinear sine amplification function of Fig.~\ref{fig:linnonlinamp} one
arrives at the harmonic spectrum in the right-hand plot of Figure \ref{fig:4notes}.
Inter-modulation distortion has generated substantial harmonic content at 375, 525, 825, and 975
Hz, thus filling in the harmonic series of a fundamental at 75 Hz.
Indeed the harmonic series continues all the way to the 21st harmonic at 1575 Hz.
Even though we didn't play the low D, we can hear it!

\section{Summary}

It's amazing how nonlinear amplification and a few trigonometry relations can unlock the creation
of new notes not played by the musician.  With distorted amplification the input can be simple --
only to notes are required to generate new notes in the harmonic spectrum, a spectrum reminiscent
of an acoustic guitar played with full bar chords.

And for those of you on a keyboard running through a distorted amplifier -- thinking about {\em Deep
Purple} again, perhaps showing our age\ldots -- remember the perfect 4th and perfect 5th intervals
are almost perfect on an equal temperament keyboard.  These intervals are going to sound great, and
in any key too.  But stay away from the intervals of a major 3rd or minor 7th.  These intervals are
out of harmonic tuning and will sound bad when distorted by an amplifier.

We hope that you have shared our amazement at the complicated combinations of frequencies generated
by distorted amplifiers in modern music. For a story that started out by singing praises for {\em
  simple frequency ratios} of small integer numbers -- the basis of harmony according to Pythagoras
-- this tale about the physics of music has surprised us with its complexity. Understanding the
physics behind our favourite songs and sounds leaves us with a deep understanding on how to make
the best sounding music with modern distorted amplifiers.

\section*{Acknowledgements}

It is a pleasure to thank the University of Adelaide for the award of a summer research scholarship
which provided the opportunity to commence the exploration of this most fascinating area of
physics.


\end{document}